\documentclass{aa}

\usepackage{appendix}
\usepackage{graphicx}
\usepackage{natbib}
\begin{document} 

\title{Distinguishing \texttt{disks} from \texttt{mergers}: tracing the kinematic asymmetries in local (U)LIRGs using \texttt{kinemetry}-based criteria}
   \author{Enrica Bellocchi\inst{1,2}, Santiago Arribas\inst{2}, Luis Colina\inst{2} }
   \institute{\small $^1$ Departamento de F\'isica Te\'orica, Universidad Aut\'onoma de Madrid, 28049 Madrid, Spain\\   
   $^2$ Centro de Astrobiolog\'ia, Departamento de Astrof\'isica, CSIC-INTA, Cra.~de Ajalvir km.4, 28850 - Torrej\'on de Ardoz  (Madrid) Spain\\
        \email{bellochie@cab.inta-csic.es; enrica.bellocchi@gmail.com}\\  
                  }

 \date{Received 16 July 2015; Accepted 26 February 2016}
 
\abstract 
{The kinematic characterization of different galaxy populations is a key observational input to distinguish between different galaxy evolutionary scenarios, since it helps to determine the number ratio of rotating disks to mergers at different cosmic epochs. Local (U)LIRGs offer a unique opportunity to study at high linear resolution and S/N extreme star forming events and compare them with those observed at high--z. }
{Our goal is to analyze in detail the kinematics of the H$\alpha$ ionized gas of a large sample of 38 local (z $<$ 0.1) (U)LIRGs (50 individual galaxies) applying kinematic criteria able to characterize the evolutionary status of these systems.}
{We obtained Very Large Telescope (VLT) VIMOS optical integral field spectroscopy (IFS) data of a sample of 38 (U)LIRGs. The `unweighted' and `weighted' \texttt{kinemetry}--based methods are used to kinematically classify our galaxies in `disk' and `merger'. We simulate our systems at z=3 to evaluate how a loss of angular resolution affects our results.}
{From the \texttt{kinemetry}-based analysis we are able classify our local (U)LIRGs in three distinct kinematic groups according to their total kinematic asymmetry values (K$_{tot}$) as derived when using the weighted (unweighted) method: 1) 25 out of 50 galaxies are kinematically classified as `disk', with a K$_{tot}\leq$ 0.16 (0.14); 2) 9 out of 50 galaxies are kinematically classified as `merger', with a K$_{tot}\geq$ 0.94 (0.66); 3) 16 out of 50 galaxies lie in the `transition region', in which disks and mergers coexist, with 0.16 (0.14) $<$ K$_{tot}$ $<$ 0.94 (0.66). 
When we apply our criteria to the high--z simulated systems, a lower total kinematic asymmetry frontier value (K$_{tot}$ $\sim$ 0.16 ($\sim$ 0.14)) is derived with respect to that found locally. The loss of angular resolution smears out the kinematic features, thus making objects to appear more kinematically regular than actually they are.}
{}


\titlerunning{Kinematics asymmetries of (U)LIRGs}
\authorrunning{Bellocchi et al.}

\maketitle

\section{Introduction}

In the standard model of hierarchical galaxy assembly mergers are the dominant source of mass accretion and growth in massive high-redshift (high-z) galaxies (i.e., \citealt{cole00, Somerv01}). In this scenario, galaxies are assumed to form at the center of the dark matter halos as the baryonic gas cools (e.g., \citealt{baugh06}), and their subsequent evolution is controlled by the merging histories of the halos containing them (e.g., \citealt{Cole94}). As derived from the observations, major merging is undoubtedly taking place at high-z (e.g., \citealt{T06, T08}). More recently, \citet{kartal12}, studying a sample of ULIRGs at z $\sim$ 2, found that the majority of the sources show signs of major mergers. These mechanisms support the hypothesis that gas-rich late-type galaxies can transform into gas-poor early-type E/S0 galaxies, as predicted using detailed simulations (\citealt{MH96, combes04, Conse06}). As a result of this framework, we expect to find galaxies characterized by complex and disturbed kinematics, such as distorted and asymmetric velocity fields, as a proof of a strong ongoing interaction.

On the other hand, hydrodynamical simulations (\citealt{Robertson06, Governato07, Governato09, Hokpins09a, Hopkins09}) have shown that the gas fraction at fusion time and the amount of dissipation in a major merger of disk galaxies is a key parameter to generate a bulge dominated/elliptical (i.e., through dry merger) or a spiral galaxy (i.e., through gas-dominated merger). Indeed, according to this scenario a disk can be reformed in the remnant when the fraction of gas at the fusion time is higher than 50\% (e.g., \citealt{Hammer09, Puech12}).

%2
In the last few years many works found that most of the high--z galaxies show as a disk-like rotating velocity field pattern, although they appear to be turbulent (i.e., \citealt{Lehn09, Burk10}) as given by their high velocity dispersion (i.e., $\sigma$ = 30 -- 100 km s$^{-1}$) and low dynamical ratio (v/$\sigma<$ 1; \citealt{genzel08, FS09, Wisn2011}). In order to explain their kinematic patterns in new models of disk formation at high--z, recent theoretical works (i.e., \citealt{keres05, dave08, genel08, dekel09_a, ceverino10}), based on semi-analytical approaches and hydrodynamical simulations, have invoked a rapid but more continuous gas accretion via {\it cold flows} and/or minor mergers, which likely play an important role in driving the mass growth of massive star forming galaxy at high--z (e.g., \citealt{Daddi07, elbaz07, noeske07}), able to supply gas directly to the center of the galaxies (i.e., \citealt{keres05, Ocv08, dekel09_a}). One of the first evidences of the `clumpy disk' picture came from the work of \citet{FS06}, observing the H$\alpha$ emission of a sample of 14 BM/BX galaxies. They confirmed the presence of a significant fraction of galaxies with rotation fields characteristic of disks, large enough to be resolved in 0.5 arcsec seeing. Then, a large portion of the strongly star-forming galaxies (SFGs) at z = 1 -- 3 does not show the disturbed kinematics expected according to the hierarchical model but is characterized by regularly rotating disks (e.g., \citealt{cresci09, Epi09, FS09, gnerucci11}). This result has suggested that even stronger star formation may be fueled by the accretion of pristine gas from the halo and by dynamical instabilities within the massive gaseous disks (\citealt{genel08, dekel09, Bouche10, cresci10}).

The discrepancies between morphological and kinematical results have increased the importance of kinematic studies, since objects photometrically irregular in broad-band {\it HST} images show `regular' kinematic maps (i.e., \citealt{Bou08, vStark08, puech10, jones10, FS11, genzel11}). Thus, the aforementioned results emphasize the crucial role of spatially- and spectrally-resolved investigations of galaxies at different redshifts, such as those based on integral field spectroscopy (IFS), in order to map their morphology and kinematics.

A useful way to figure out which is the dominant scenario that drives the galaxy evolution at different cosmic epochs is to estimate the number ratio of {\it (rotating) disks} to {\it mergers} (i.e., disk/merger fraction). Some discrepancies can raise in classifying several kinds of galaxies in {\it disks/mergers} when applying different techniques. The first and most widely used technique was the visual morphological classification (e.g., \citealt{dasyra08, kartal10, Zamo11}); then other classification methods, as the estimate of the asymmetry and clumpiness parameters (e.g., \citealt{Conse03}) and the use of the Gini-M$_{20}$ plane (e.g., \citealt{Lotz08}), have been considered as well. In the last decade, measures of galaxy kinematics (e.g., \citealt{genzel08, FS06, FS09}) have increased their importance: in order to kinematically classify the systems, a visual (kinematic) classification has been applied at intermediate redshift (i.e., 0.4 $\leq$ z $\leq$ 0.7) by \citet{Flores06} and \citet{Yang08} and locally by \citet{YO2013}  to investigate the properties of the velocity fields of galaxies observed with IFS.

%Recently, several authors (e.g., \citealt{S08, Gon10, swin12}, alaghband) used the {\it kinemetry} methodology (\citealt{K06}; see Fig.~\ref{shap}) which allows to quantify the kinematic asymmetries of several galaxy samples located at different redshifts (mainly at high--z) from analyzing their velocity fields and velocity dispersion maps in order to distinguish and characterize {\it merging} and {\it non-merging (disk)} systems. Taking advantage of this methodology, \citet{jesseit07} and \citet{kron07} investigated the distortions in the velocity fields of simulated interacting {\it disk-merger} galaxies at different redshifts, between z = 0 - 1. \citet{jesseit07} tried to understand the possible formation of ellipticals by merger of disks. They found that disk merger remnant generated by equal mass (1:1) merger are rounder than 1:3 remnant. Then, the non-rotating subset of the representative SAURON sample (i.e., \citealt{Emsel04}) of local galaxies agrees with 1:1 merger simulation, while the rotating subset can be reproduced by the 3:1 merger remnant.  On the other hand, \citet{kron07} also investigated resolution effects simulating distorted velocity fields at high--z, founding that distortions are still visible at intermediate redshift for large (Milky Way type) galaxies while for small (disk scale length $\sim$ 2 kpc) galaxies strong distortions are not visible in the velocity fields.

The kinematic characterization of different galaxy populations (\citealt{Glazebrook13} review) is a key observational input to distinguish between different galaxy evolutionary scenarios, since it helps us to determine the number ratio of rotating disks to mergers at different cosmic epochs. This provides a way of constraining the relative role of major mergers and steady cool gas accretion in shaping galaxies, which remains a topic of discussion (e.g. \citealt{genzel01, T08, dekel09,  FS09, lemoine09, lemoine10, Bou11, Epi11}).To this aim several authors have already analyzed the velocity fields and velocity dispersion maps of different galaxy samples (e.g., Lyman break analogs LBAs, Sub-mm galaxies SMGs, (U)LIRGs, H$\alpha$ emitters, high--z simulated SINGS spiral galaxies) using the \texttt{kinemetry} methodology\footnote{\texttt{Kinemetry} is a tool able to quantify kinematic asymmetries in the velocity field and velocity dispersion maps of the systems with respect to those characterizing an ideal rotating disk. It will be described in the Sect. 3.} (\citealt{K06}, hereafter, K06) with the aim of discerning merging and non-merging systems on the base of their kinematic properties (e.g., \citealt{S08, Gon10, AZ12, YO2012, swin12, Hung15}).

%1
LIRG (L$_{IR}$ = [8 - 1000 $\mu$m] = 10$^{11}$ - 10$^{12}$ L$_\odot$) and ULIRG (ULIRGs, L$_{IR}$ $>$ 10$^{12}$ L$_\odot$) galaxy populations are particularly relevant to the study of galaxy evolution since, although rare in the local universe,  they are far more numerous at high--z and responsible for a significant fraction of previous star formation prior to redshift z $\sim$ 1 (e.g., \citealt{LF05, PG05, PPG08}). Several authors have suggested that high--z LIRGs are scaled-up versions of low--z LIRGs (e.g., \citealt{pope06, papov07, elbaz10, elbaz11, nordon10, nordon12, TAK10}), finding that in the local universe (U)LIRGs cover a similar SFR range than normal high--z SFGs (e.g., \citealt{wuyts11}). Therefore, low--z (U)LIRGs offer a unique opportunity to study at high linear resolution and S/N extreme star forming events and compare them with those observed at high--z. 

%As done in \citet{YO2012} we apply the {\it kinemetry}-based criteria to the observed high S/N and simulated `high--z' kinematic maps in order to distinguish in between {\it disks} and {\it mergers} systems. 

%2
In this paper we present the results from applying the \texttt{kinemetry} method to a large sample of 38 local (U)LIRG systems (51 individual galaxies) observed with the VIMOS/VLT integral field unit (IFU). The same approach as in  \citet{YO2012} will be taken into account considering both locally observed and high--z simulated (U)LIRGs systems. Thus, this study will allow us to constrain the disk/merger fraction in the local universe as well as to compare such ratio with that derived for high--z populations.

The paper is structured as follows. In Sect. 2, we present the sample giving details about the observations, data reductions, line fitting and map construction. Sect. 3 is devoted to the description of the \texttt{kinemetry} analysis and its potential in distinguishing {\it disks} from {\it mergers} when applying two different methods (i.e., \citealt{S08, YO2012}, hereafter, S08 and B12) to a sample of local and high--z simulated galaxies. Finally, the main results and conclusions are summarized in Sect. 4. Throughout the paper we will consider H$_0$ = 70 km s$^{-1}$ Mpc$^{-1}$, $\Omega_M$ = 0.3 and $\Omega_\Lambda$ = 0.7.

\section{Observations, data reduction and data analysis}

\subsection{The sample and the morphological class}
\label{morph_class}

\begin{figure}[htbp]
\begin{center}
\includegraphics[width=0.45\textwidth]{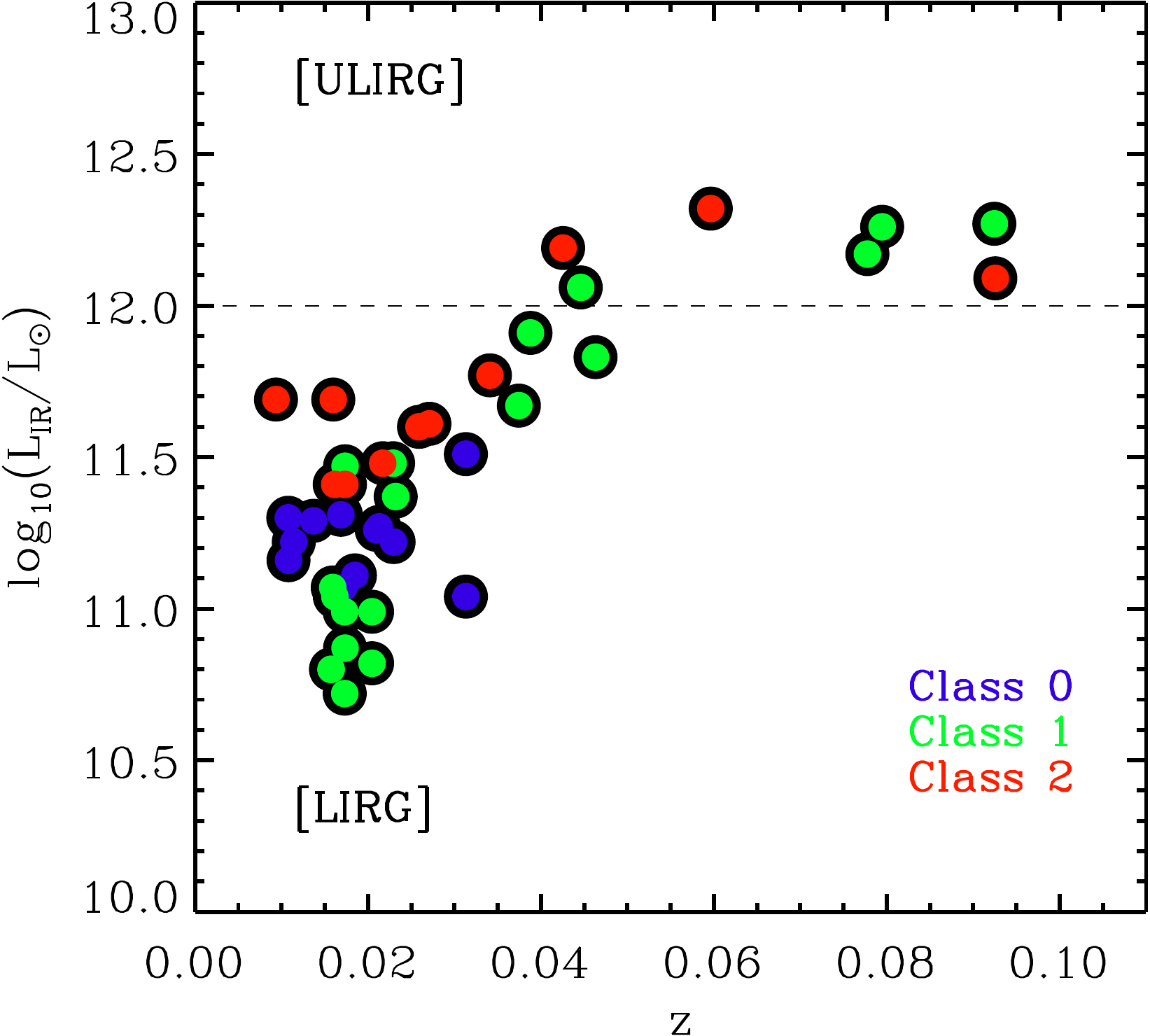}
\caption{Distribution of the VIMOS sample in the infrared luminosity -- redshift plane. The different colors represent the three stages of interactions: in blue are shown the `Class 0'  single isolated objects, in green the `Class 1' interacting galaxies in a pre-coalescence phase and in red the `Class 2'  objects representing single systems with evidence of having suffered a merger (post-coalescence phase). The horizontal dashed line separates the LIRG and ULIRG domains.}
\label{fig:uncomplete}
\end{center}
\end{figure}

The (U)LIRG sample analyzed in this work is the same than that analyzed in \citet{YO2013} (hereafter, B13) in which the 2D kinematic properties of the ionized gas (H$\alpha$) are discussed. To summarize, it contains a total of 38 (U)LIRGs systems (51 individual galaxies) of the southern hemisphere drawn from the Revised Bright Galaxy Sample (RBGS, \citealt{sanders03}). Of these systems 31 are LIRGs (i.e., $<$L$_{IR}>$ = 2.9 $\times$10$^{11}$ L$_\odot$) with a mean redshift of 0.024 (corresponding to D $\sim$ 100 Mpc), and the remaining seven are ULIRGs (i.e., $<$L$_{IR}>$ =1.6 $\times$ 10$^{12}$ L$_\odot$) with a mean redshift of 0.069 (D $\sim$ 300 Mpc; see Tab. \ref{table_sample} and \citet{A08}, hereafter A08, for details). This sample thus includes a good representation of the relatively less studied LIRG luminosity range. It also encompasses a wide variety of morphological types, suggesting different dynamical phases (isolated spirals, interacting galaxies, and ongoing- and post-mergers), and nuclear excitations (HII, Seyfert, and LINER). Eleven out of 51 galaxies show evidence in their optical nuclear spectra of hosting an AGN, showing high  [NII]/H$\alpha$ values and/or broad H$\alpha$ emission lines (e.g., IRAS F07027-6011N, IRAS F05189-2524, IRAS F12596-1529, IRAS F21453-3511; see \citealt{MI10, A12, A14} hereafter, MI10, A12, A14, respectively). Most of these objects (46 out of 51) show outflows of ionized gas, studied in A14, while a smaller fraction (22 out of 40) show outflows of neutral gas, studied in \citet{Cazzoli14, Cazzoli16}. The sample is not complete either in luminosity or in distance. However, it covers well the relevant luminosity range and is representative of the different morphologies within the (U)LIRG phenomenon (see Fig. \ref{fig:uncomplete}).

The morphological class was derived using Digital Sky Survey (DSS) ground-based images and, when available, additional archival Hubble Space Telescope (HST) images. Except for one case, the morphological classification according to the DSS images is in good agreement with that derived by using the HST images\footnote{For 19 out of 38 of the systems in our sample DSS and HST images are available. Only for IRAS F06206-6315 the HST image reveals more features than those shown in the DSS image, morphologically classifying this galaxy as type 1 vs. type 2/0 when using, respectively, HST and DSS images.}. The sources have been morphologically classified following a simplified version of the scheme proposed by \citet{V02}, with three main classes instead of five (see \citealt{RZ11}, hereafter RZ11, and references therein for further details). We refer to the images published in B13 (in their Appendix A) in which the H$\alpha$ kinematic maps are shown for each galaxy and to their DSS/HST images published in RZ11. Briefly, we remind the three morphological classes defined as follows:

\vspace{2mm}

\begin{itemize}
\item Class 0: objects that appear to be single isolated galaxies, with a relatively symmetric disk morphology and without evidence for strong past or ongoing interaction (hereafter, $disk$).
\item Class 1: objects in a pre-coalescence phase with two well-differentiated nuclei separated a projected distance $>$ 1.5 kpc. For these objects, it is still possible to identify the individual merging galaxies and, in some cases, their corresponding tidal structures due to the interaction (hereafter, $interacting$). 
\item Class 2: objects with two nuclei separated a projected distance $\leq$ 1.5 kpc or a single nucleus with a relatively asymmetric morphology suggesting a post-coalescence merging phase (hereafter, $merger$).
\end{itemize}

In Tab. \ref{table_sample} we present the main properties of the sample. In some cases the properties of individual galaxies in multiple systems could be inferred separately and were therefore treated individually (see B13 for details). 

%As part of the analysis, additional galaxies with available IFS data have been included to increase the number of systems in the ULIRG luminosity range. These data (\citealt{Co05, GM07, GM09}) correspond to our own observations taken with INTEGRAL/WHT.

%________________________________________________________________
\subsection{Observations and data reduction}

The observations have been described in detail in previous papers (i.e., A08, MI10, RZ11, B13). In brief, they were carried out  using the Integral Field Unit of VIMOS (\citealt{Lfevre03}), at the Very Large Telescope (VLT), covering the spectral range $(5250-7400)$ \AA\ with the high resolution  grating GG435 (`HR-orange' mode) and a mean spectral resolution of 3470 (dispersion of 0.62 \AA\ pix$^{-1}$). The effective field of view (FoV) in this configuration is 29.5\arcsec $\times$ 29.5\arcsec, with a spaxel scale of 0.67\arcsec per fiber (i.e., 1936 spectra are obtained simultaneously from a 44 $\times$ 44 fibers array).

The VIMOS data are reduced with a combination of the pipeline {\it Esorex} (versions 3.5.1 and 3.6.5) included in the pipeline provided by ESO, and different customized IDL and IRAF scripts. The basic data reduction (i.e., bias subtraction, flat field correction, spectra tracing and extraction, correction of fiber and pixel transmission and relative flux calibration) is performed using the {\it Esorex} pipeline. The four quadrants per pointing are reduced individually and then combined into a single data cube. Then, the four independent dithered pointing positions are combined together to end up with the final `super-cube', containing 44 $\times$ 44 spaxels for each object (i.e., 1936 spectra). For the wavelength calibration description we refer to A08 and RZ11. 

%The wavelength calibration, the instrumental profile, and fiber-to-fiber transmission correction are checked for all the galaxies using the [OI]$\lambda$6300.3 \AA\ sky line. We fitted this line to a single Gaussian for all spectra of each individual source, obtaining a central wavelength and a FWHM for each object. The mean values, when all the sources are considered, are respectively (6300.29 $\pm$ 0.07) \AA\ and (1.80 $\pm$ 0.07) \AA\, and give a good estimate of our uncertainties due to calibration.  A detailed description of the data reduction process is discussed in Paper I and III.

\subsection{Data analysis}
\label{reduc}

The observed H$\alpha$ and [NII]$\lambda\lambda$6548, 6583 \AA\ emission lines of the individual spectra are fitted to Gaussian profiles using an IDL routine (i.e., MPFITEXPR, implemented by C. B. Markwardt). This algorithm derives the best set of lines that match the available data. In case of adjusting multiple lines, the line flux ratios and wavelengths of the different lines are fixed according to the atomic physics. The widths are constrained to be equal for all the lines and greater than the instrumental contribution ($\sigma_{INS}$). The results of the fit have been presented in the Appendix A in B13, in which the H$\alpha$ maps are shown (i.e., flux intensity, velocity field and velocity dispersion maps). 

As largely described in B13, a \texttt{narrow} (or \texttt{systemic}) and \texttt{broad} components\footnote{The distinction between \texttt{narrow} and \texttt{broad} components has been done according to their line widths.} have been identified  in most of the systems to properly fit the spectra. In this analysis we will focus on the kinematic maps of the \texttt{systemic} component, assumed to be the narrow component of the emission line, since the spatial distribution and kinematic properties of this component represent those of the entire galaxy.

\subsection{Simulated high--z observations: the resolution effects}
\label{SIMS}

In order to investigate how a decreasing angular resolution affects our results, we simulate observations at z = 3 with a typical pixel scale of 0.1$^{\prime\prime}$ (the same pixel scale as the IFU NIRSpec/JWST) as done in B12, just considering resolution effects. Since the angular distance evolves less than 10\% in the redshift range z = 2 -- 3, our simulated observations at z = 3 are relevant for a direct comparison to observations at z $\sim$ 2.

The `simulated' FoV of the maps ranges between $\sim$1$^{\prime\prime}$$\times$1$^{\prime\prime}$ up to 5$^{\prime\prime}$$\times$5$^{\prime\prime}$ with the scale of about 7.7 kpc arcsec$^{-1}$ assuming the $\Lambda$CDM cosmology considered in this work.

\begin{table*}
\vspace{-3.3mm}
\centering
\caption{General properties of the (U)LIRG sample.}
\label{table_sample}
\begin{scriptsize}
\vspace{-3.5mm}
\begin{tabular}{l c c c c c c c } \\
\hline\hline\noalign{\smallskip}  
ID1{} &  ID2{} &   $z$ &  D&  scale{} & log L$_{IR}$  {}& Class    &  Notes   \\
IRAS & Other &  &  (Mpc) &  (pc/$^{\prime\prime}$)     & (L$_{\odot}$)    &        &    \\ 
(1) & (2) & (3) & (4) & (5) & (6) & (7) & (8)   \\
\hline\noalign{\smallskip} 	
F01159--4443 S  &   ESO 244--G012 	&  0.022903 &  99.8 		&  462  & -   					&  1   & a,d \\ 
F01159--4443 N &   ESO 244--G012 	&  0.022903 &  99.8 		&  462  &11.48  				&  1   & a,d \\ 
\hline\noalign{\smallskip} 	
F01341--3735 S &  ESO 297--G012 		&  0.017305 & 75.1  		&  352  & 10.72 				& 1   & a,d \\
F01341--3735 N  &  ESO 297--G011 	&  0.017305 & 75.1  		&  352  & 10.99 				& 1   & a,d \\
\hline\noalign{\smallskip} 	
F04315--0840  &  NGC 1614 			& 0.015983  & 69.1  		& 325   & 11.69  				& 2 	 &  \\
\hline\noalign{\smallskip} 	
F05189--2524  &   					& 0.042563  & 188.2  		&  839  & 12.19  				& 2  &\\
\hline\noalign{\smallskip} 	
F06035--7102  &  						& 0.079465  &  360.7 		& 1501 & 12.26  				& 1 	 & \\
\hline\noalign{\smallskip} 	
F06076--2139  S &  					& 0.037446  & 165  		&  743  & -		 			& 1  &a,d \\
F06076--2139  N &  					& 0.037446  & 165  		&  743  & 11.67  			& 1  &a,d \\
\hline\noalign{\smallskip} 	
F06206--6315  &  						&0.092441   & 423.3 		& 1720 &  12.27  				&1 	 &\\
\hline\noalign{\smallskip} 	
F06259--4780 S & ESO 255--IG007 		& 0.038790  & 171.1  		& 769   &  -					& 1 	 & b,d\\
F06259--4780 C & ESO 255--IG007 		& 0.038790  & 171.1  		& 769   &  -		 			& 1 	 & b,d\\
F06259--4780 N & ESO 255--IG007 		& 0.038790  & 171.1  		& 769   &  11.91   			& 1 	 & b,d\\
\hline\noalign{\smallskip} 	
F06295--1735  & ESO 557--G002 		& 0.021298  & 92.7  		&  431  &  11.27  				& 0   &\\
\hline\noalign{\smallskip} 	
F06592--6313  &  						& 0.022956  & 100 		& 464  & 11.22 				& 0 	 &\\
\hline\noalign{\smallskip} 	
F07027--6011 S &  AM 0702--601 		& 0.031322 & 137.4 		& 626  & 11.51				& 0 &a,d \\
F07027--6011 N &  AM 0702--601 		& 0.031322 & 137.4 		& 626  & 11.04				& 0 &a,d \\
\hline\noalign{\smallskip} 	
F07160--6215 & NGC 2369 			& 0.010807 & 46.7 		& 221  & 11.16 				& 0  & \\
\hline\noalign{\smallskip} 	
08355--4944   &  						& 0.025898 & 113.1 		& 521  & 11.60  				& 2  &\\
\hline\noalign{\smallskip} 	
08424--3130 S  & ESO 432--IG006 		& 0.016165 & 70.1 		&329   &11.04 				& 1 &a,d \\
08424--3130 N  & ESO 432--IG006 		& 0.016165 & 70.1 		&329   &-					& 1 	  &a,d \\
\hline\noalign{\smallskip} 	
F08520--6850 E & ESO 60--IG016 		& 0.046315 & 205.4 		& 909  & 11.83 				& 1 	 &\\
F08520--6850 W & ESO 60--IG016 		& 0.046315 & 205.4 		& 909  & 11.83 				& 1 	 &\\ 
\hline\noalign{\smallskip} 	
09022--3615   &  						& 0.059641 & 267 		&1153 &12.32 				& 2 	 &\\
\hline\noalign{\smallskip} 	
F09437+0317 S & IC 563  			& 0.020467 & 89 			& 415  & 10.82				& 1 (0)& a,c,d \\
F09437+0317N & IC 564 				& 0.020467 & 89 			& 415  & 10.99				& 1 (0)& a,c,d \\
\hline\noalign{\smallskip} 	
F10015--0614 & NGC 3110 			& 0.016858 & 73.1 		&343   & 11.31 				& 0 	 &\\
\hline\noalign{\smallskip} 	
F10038--3338 &  ESO 374--IG032		& 0.034100 & 149.9 		& 679  & 11.77	 			& 2 	 &\\ 
\hline\noalign{\smallskip} 	
F10257--4339 & NGC 3256			& 0.009354 & 40.4 		& 192  & 11.69 				& 2 	 &\\
\hline\noalign{\smallskip} 	
F10409-4556 & ESO 264-G036 		& 0.021011 & 91.4 		& 425  & 11.26 				& 0 	 &\\
\hline\noalign{\smallskip} 	
F10567--4310 & ESO 264--G057 		& 0.017199 &  74.6  		&  350  & 11.07  				& 0 	 &    \\
\hline\noalign{\smallskip} 	
F11255--4120 &  ESO 319--G022 		& 0.016351 & 70.9 		& 333  & 11.04  				& 0 	 &     \\
\hline\noalign{\smallskip} 	
F11506--3851 & ESO 320--G030		& 0.010781 & 46.6 		& 221  & 11.30 				& 0 	 &\\
\hline\noalign{\smallskip} 	
F12043--3140 S & ESO 440--IG058 		& 0.023203 & 101.1 		& 468  & 11.37 				& 1	 &a,d \\ 
F12043--3140 N & ESO 440--IG058 		& 0.023203 & 101.1 		& 468  & -		 			& 1	 &a,d \\ 
\hline\noalign{\smallskip} 	
F12115--4656 & ESO 267--G030 		& 0.018489 & 80.3 		& 375  &11.11				& 0 &\\ 
\hline\noalign{\smallskip} 	
12116--5615   & 						&  0.027102 & 118.5 		& 545  &11.61 				& 2 (0) &\\
\hline\noalign{\smallskip} 	
F12596--1529 & MCG 02--33--098 		& 0.015921 & 69.0 		& 324  &11.07 				& 1 	 & \\
\hline\noalign{\smallskip} 	
F13001--2339 & ESO 507--G070 		& 0.021702 & 94.5 		& 439  & 11.48 				& 2 (0/1)  &  \\
\hline\noalign{\smallskip} 	
F13229--2934 & NGC 5135 			&  0.013693 & 59.3		&  280 & 11.29 				& 0 	 & \\
\hline\noalign{\smallskip} 	
F14544--4255 E & IC 4518 			&  0.015728 & 68.2 		&  320 & 10.80 				& 1 &a,d \\
F14544--4255 W & IC 4518 			&  0.015728 & 68.2 		&  320 & 10.80 				& 1  &a,d \\
\hline\noalign{\smallskip} 	
F17138--1017 & 						&  0.017335 & 75.2 		& 352  & 11.41				& 2 (0) & \\
\hline\noalign{\smallskip} 	
F18093--5744 S &  IC 4689			&  0.017345 & 75.3 		& 353  & - 					& 1  &b,d	\\
F18093--5744 C &  IC 4686			&  0.017345 & 75.3 		& 353  & 10.87				& 1  &b,d	\\
F18093--5744 N &  IC 4687			&  0.017345 & 75.3 		& 353  & 11.47				& 1  &b,d	\\
\hline\noalign{\smallskip} 	
F21130--4446 &  						&  0.092554 & 423.9 		& 1722 & 12.09 				& 2 	 &\\
\hline\noalign{\smallskip} 	
F21453--3511 &  NGC 7130 			& 0.016151  & 70.0 		& 329   & 11.41  				& 2  &    \\
\hline\noalign{\smallskip} 	
F22132--3705 &  IC 5179 				& 0.011415  & 49.3 		& 234   & 11.22  				& 0 	 &   \\
\hline\noalign{\smallskip} 	
F22491--1808 &  						& 0.077760  & 352.5 		& 1471 & 12.17 				& 1 &\\
\hline\noalign{\smallskip} 	
F23128--5919 &  AM 2312--591			& 0.044601  & 197.5 		& 878   & 12.06   				& 1  &  \\
\hline\hline
\end{tabular}
\end{scriptsize}
 \begin{minipage}{18cm}
   \vskip0.01cm\hskip0.0cm
\footnotesize
\tablefoot
{Col (1): Object designation in the Infrared Astronomical Satellite (IRAS) Faint Source Catalog (FSC). 
Col (2): Other identification. 
Col (3): Redshift from the NASA Extragalactic Database (NED). 
Col (4): Luminosity distance assuming a $\Lambda$DCM cosmology with H$_0$ = 70 km s$^{-1}$ Mpc$^{-1}$, $\Omega_M$ = 0.3, and $\Omega_\Lambda$ = 0.7, using the E. L. Wright Cosmology calculator, which is based on the prescription given by \citet{Wri06}. 
Col (5): Scale. 
Col (6): Infrared luminosity (L$_{IR}$= L(8-1000) $\mu$m) in units of solar bolometric luminosity, calculated using the fluxes in the four IRAS bands as given in \citet{sanders03} when available. Otherwise, the standard prescription given in \citet{SM96} with the values in the IRAS Point and Faint Source catalogs was used. 
Col (7): Morphological class defined as follows: 0 identifies {\it isolated} objects, 1 {\it pre-coalescence} systems, and 2 stands for {\it merger} objects. For those objects for which the morphological classification is uncertain, the various possible classes are shown in the table with the preferred morphological classification indicated in the first place and the alternative classification within brackets (see text for further details). 
%Col (8): Kinematical classification as derived in Paper IV. 
Col (8): Notes with the following code: 
(a) System composed of two galaxies. 
(b) System composed of three galaxies. 
(c) There are two VIMOS pointings for the northern source. 
(d) Interacting system (i.e., see notes $a$ and $b$) for which the total infrared luminosity L$_{IR}$ could be approximately assigned among the members of the system according to the MIPS/Spitzer photometry. For further details see B13.}
\end{minipage}
\end{table*}

\section{\texttt{Kinemetry} analysis}

We investigate the power of the \texttt{kinemetry} methodology in studying the kinematic asymmetries in (U)LIRG systems. In particular, the same approach described in B12 (i.e., using the S08 and B12 criteria, hereafter, `unweighted' and `weighted' methods respectively) is applied to i) the entire observed local sample\footnote{The final total number of galaxies analyzed in this work with \texttt{kinemetry} is 50 instead of 51, since the galaxy F08424-3130 N is located in the edge of the VIMOS FoV and then excluded from this analysis.} and then to ii) the simulated high--z kinematic maps. In this section the results derived from applying these two methodologies are discussed.

\subsection{The method}

The \texttt{kinemetry} method comprises a decomposition of the moment maps into Fourier components using ellipses. We briefly describe the main steps presented in K06 to achieve a clearer understanding of this analysis.

The Fourier analysis is the most straightforward way to characterize any periodic phenomenon: the periodicity of a kinematic moment can easily be seen by expressing the moment in polar coordinates where K (x, y) $\rightarrow$ K (r, $\psi$). The map K(r, $\psi$) can be expanded as follows to a finite number (N+1) of harmonic terms (frequencies)

\begin{equation}
 K(r, \psi) = A_{0}(r) + \sum _{n=1}^N A_{n}(r) \hspace{1mm}sin(n \cdot \psi) + B_{n}(r)\hspace{1mm} cos(n \cdot\psi) 
\end{equation} 

where $\psi$ is the azimuthal angle in the plane of the galaxy (measured from the major axis) and $r$ is the radius of a generic ellipse. The amplitude and the phase coefficients ($k_{n}$, $\phi_{n}$) are easily calculated from the $A_{n}$, $B_{n}$ coefficients as $k_{n} = \sqrt{A_{n}^{2} + B_{n}^{2}}$ \hspace{1mm} and $\phi_{n} = arctan \left(\frac{A_{n}}{B_{n}}\right)$. 

For an \textit{ideal rotating disk} the $B_1$ term dominates the velocity profile, representing the {\it circular} velocity in each ring $r$ while the $A_0$ term dominates the velocity dispersion profile giving gives the {\it systemic velocity} of each ring. Thus, higher order terms ($A_{n}$, $B_{n}$) indicate deviations from symmetry. In the \texttt{kinemetry} analysis we assume for each ellipse a covering factor = 0.7, a position angle ($\Gamma$) and a flattening ({\it q}) free to vary and the peak of the VIMOS continuum emission as the center of the ellipse. 

The covering parameter represents the minimum percentage of the points along an ellipse needed to start the analysis. In our case 0.7 means that if fewer than 70\% of the points along an ellipse are not covered by data the program stops. The position angle of the velocity field ($\Gamma$) identifies the angle where the velocity field peaks with respect to the North coordinate. The flattening ($q$) is defined as the ratio of the semi-minor (b) to the semi-major (a) axes of the ellipse, i.e., q= b/a. When let free to vary it allows us to consider general cases, such as tilted/wrapped disks. For further details on these assumptions see B12.

\subsection{The kinematic criteria: the `unweighted' and `weighted' methods}
 
We have considered different kinematic criteria with the aim of distinguishing systems who have suffered recent major merger events (i.e., \textit{mergers}) and those without any signs of interacting or merger activity (i.e., \textit{disks}). 

As a first approach we apply the S08 method (`unweighted' method) where the kinematic asymmetries of the velocity field and velocity dispersion maps are defined, respectively, as follows:

\begin{equation}
\hspace{1cm} v_{asym} =   \left\langle \frac{ k_{avg, v} }  {B_{1, v} }   \right\rangle_r  ,\hspace{1cm} {\sigma_{asym} =\left\langle \frac{k_{avg, \sigma}}{B_{1, v}}\right\rangle _r}
\end{equation}
\vskip3mm

where $k_{avg, v} =(k_{2, v} + k_{3, v} + k_{4, v} + k_{5, v})/4$  and $k_{avg, \sigma} =(k_{1, \sigma} + k_{2, \sigma} + k_{3, \sigma} + k_{4, \sigma} + k_{5, \sigma})/5$.

On the other hand, the method presented in B12 (`weighted' method) has been applied as well. This revised method is based on the results that indicate that in a post-coalescence merger the inner regions are dominated by rotation while the outer parts retain larger kinematic asymmetries (e.g., \citealt{kron07}). With this criterion the importance of the kinematic asymmetries at larger radii is enhanced. Indeed, instead of simply averaging the asymmetries over all radii (as in S08), they are weighted according to the number of data points used in their determination. The number of data points is to first approximation proportional to the circumference of the ellipse; the circumferences (C$_n$) of the ellipses are computed as shown in Eq. 5 in B12. The asymmetries found in the outer ellipses contribute more significantly to the average when deriving v$_{asym}^w$ and $\sigma_{asym}^w$. We remind the final formulas to compute the weighted velocity and velocity dispersion asymmetries:

\begin{equation}
\hspace{5mm}v_{asym}^w = \sum_{n=1}^N  \left (\frac{ k_{avg, n}^v}{B_{1, n}^v } \cdot C_n \right )\cdot \frac{1}{\sum_{n=1}^N C_n } ,
\label{eq:v_weighted}
\end{equation}

\begin{equation}
\hspace{5mm}\sigma_{asym}^w = \sum_{n=1}^N  \left ( \frac{ k_{avg, n}^\sigma}{B_{1, n}^v } \cdot C_n \right )\cdot \frac{1}{\sum_{n=1}^N C_n} , 
\label{eq:s_weighted}
\end{equation} 
\vskip3mm

where {\it N} is the total number of radii considered, C$_n$ the value of the circumference for a given ellipse, the different k$_n$ (k$_n^ v$ and k$_n^ \sigma$) are the deviations concerning respectively the velocity field and velocity dispersion maps, and $B_{1}^ v$ is the rotational curves.

This is the first attempt in applying the \texttt{kinemetry} method along with kinematic criteria to a large sample local (z $<$ 0.1) SFGs. This is crucial to understand what is the fraction of {\it disks} and {\it mergers} locally in such systems allowing us to compare such a ratio with those derived for high--z SFGs.

\subsection{Morphological definition of {\it disks} and {\it mergers} }
\label{morph_res}

As previously described in Sect. 2, our sample consists of 50 individual galaxies covering a large L$_{IR}$ range and encompasses a wide variety of morphologies which allow us to discuss the kinematic asymmetries as a function of the galaxy properties (e.g., infrared luminosity L$_{IR}$, morphological class).

In order to better interpret and discuss our data we recall the \citet{V02} classification, according to which the Interacting (type 1) galaxies can be sub-classified as {\it Wide-- or Close--Interacting} according to their projected nuclear separation. The presence (or not) of tidal tails and/or other structures interconnecting the nuclei is considered using DSS continuum maps and, when available, HST images since it could help us to better distinguish their structure. In particular, if the nuclear projected separation is $>$ 10 kpc, the emission of the two galaxies can be well separated in the VIMOS maps, and there is no presence of tidal tails and/or other structures between the nuclei in their continuum (DSS, HST) maps, the sources are considered as {\it Wide--Interacting pairs} (or {\it paired--disks}). A few systems (i.e., IRAS F06035--7102, IRAS F06206--6315, IRAS F12596--1529, IRAS F22491--1808, IRAS F23128--5919) are classified as {\it Close--Interacting} pairs (or {\it ongoing mergers}), since their projected nuclear separation is smaller than 10 kpc (but larger than 1.5 kpc, at which the limit for the coalescence phase is defined), their individual contributions cannot be disentangled in the VIMOS maps\footnote{The galaxy IRAS F08520--6850 is considered as {\it wide--pair} because the two galaxies can be well separated in the VIMOS maps although their nuclear separation is slightly smaller than 10 kpc.} and they also show the presence of tidal tail structures in their continuum maps. These galaxies have a common envelope, and they are likely in a more advanced merger phase than the wide-interacting pairs. Therefore, we have distinguished four dynamical phases, where the first two are referred as `disk', while the second two as `mergers' (case I): 

%\vspace{-1cm}

\begin{figure}[htbp]
\centering
\includegraphics[width=0.26\textwidth, angle=-90]{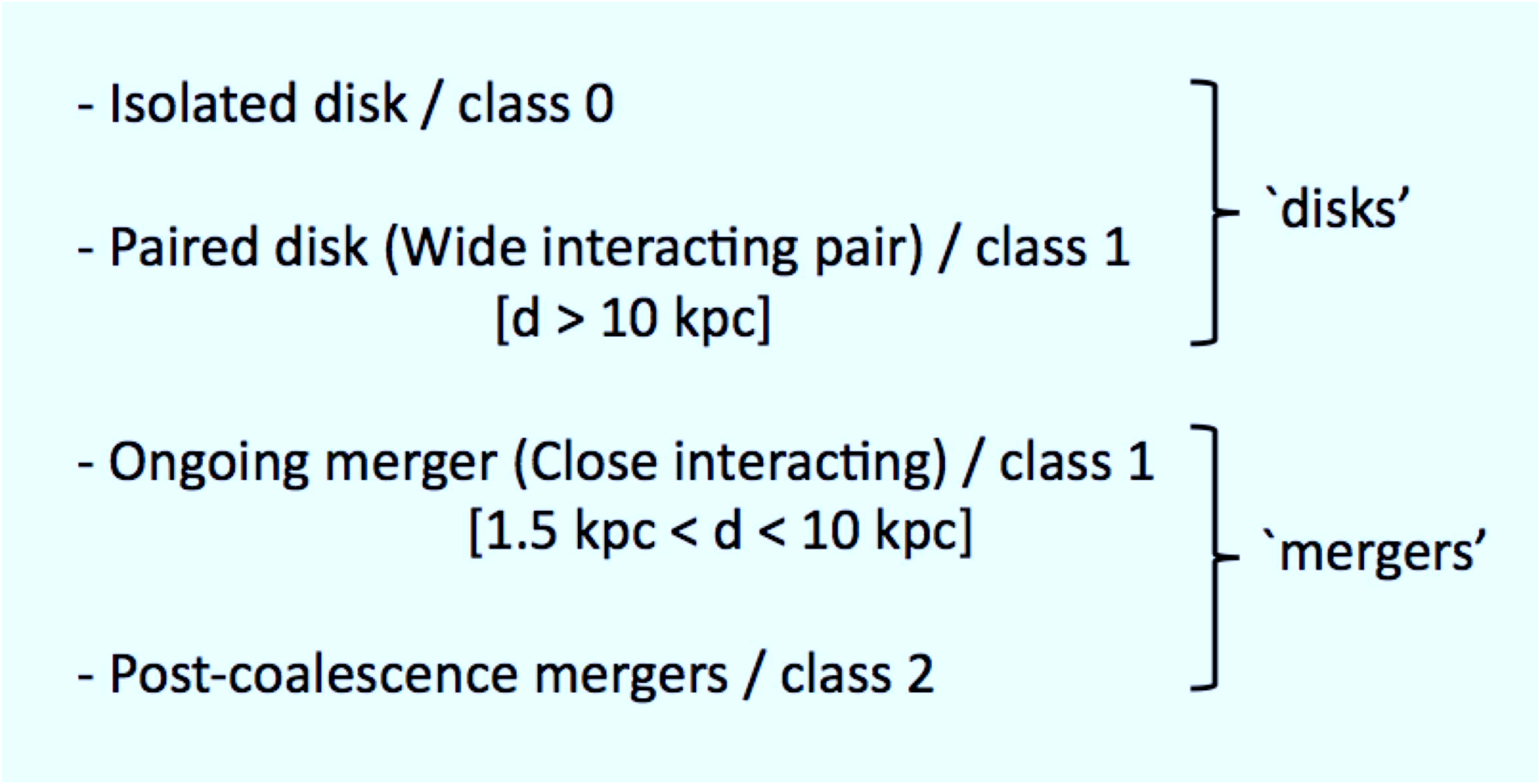}
\end{figure}

%\vspace{-2cm}

We propose such a simplified morphological (`disk--merger') classification to characterize in some way the evolutive status of the galaxies in our sample, which show a large variety of dynamical phases. In particular, as `merger' we refer to those systems where the interacting galaxies are close enough to share a common envelope as well as similar kinematics (which suggests they are probably going to merge), thus considering them as a single system. As `disk' we include the `pure' (isolated) disk galaxies as well as the `paired' galaxies which, for definition, do not share a common envelope as due to their large projected separation. In this case, we cannot surely conclude that these galaxies are going to merge and, if they do, it will be in a more advanced phase.

A similar morphological classification for the class 1 (interacting) systems has been considered in A14, in which the systems have been distinguished as `total system' or `individual object'.

However, for the isolated disks (class 0) and post--coalescence mergers (class 2) a clearer dynamical status can be inferred with respect to that derived for the interacting systems (class 1), as a result of their diversified interaction stages, something in between the two aforementioned classes. To this respect, we also discuss the \texttt{kinemetry} results derived in the case where only `isolated disks' are considered as `true disks' (case II). In this case, we exclude the `paired disk' galaxies from such a group as due to their supposed different dynamical status (i.e., interacting) with respect to the isolated disks. However, following our `merger' definition, these `paired disks' are also kept out from the `true merger' group.

\subsection{Kinematic distinction between {\it disks} and {\it mergers}}
\label{distinction}

Similarly to S08 (`unweighted' method) and what done in B12 (`weighted' method) we quantify the total kinematic asymmetry degree K$_{tot}$ of each galaxy as the combination of both the kinematic asymmetry contributions of the velocity field (v$_{asym}$) and velocity dispersion ($\sigma_{asym}$) maps, i.e., K$_{tot}$ = $\sqrt{(\sigma_{asym}^2 + v_{asym}^2)}$ (see Eqs.~2, 3, 4). We find that the ongoing merging systems have the largest kinematic asymmetry (K$_{tot}$), while (isolated and paired) disks and post-coalescence mergers are characterized by lower K$_{tot}$ values. In a similar way, ULIRGs show higher K$_{tot}$ with respect to LIRGs. The mean (median) K$_{tot}$ values for the different groups are shown in Tab. \ref{kinekine}.

\begin{table*}[htbp]
\centering
%\begin{minipage}[h]{10.5cm}
\caption{Mean (median) K$_{tot}$ asymmetry values of the (U)LIRG sample.}
\label{kinekine}
\begin{small}
\begin{tabular}{l ccr}
\hline\hline\noalign{\smallskip}
Systems & &K$_{tot}$ & \# objects \\
{\smallskip} 
 (1) & &(2) &(3) \\
\hline\noalign{\smallskip}
class 0 (Isolated disk)	&	& 0.13 $\pm$ 0.03 (0.11) &  13/50\\
class 1 (Paired disk) &	&  0.14 $\pm$ 0.03 (0.10) &  21/50\\
class 1 (Ongoing merger) 	&&  1.9 $\pm$ 0.47 (1.95) &  5/50\\
class 1 (Pair disk \& Ongoing merger)	&& 0.48 $\pm$ 0.16 (0.12) & 26/50\\
class 2 (Post-coalescence merger)	&& 0.64 $\pm$ 0.22 (0.29) & 11/50 \\
\hline\noalign{\smallskip}
LIRGs  		&&  0.27 $\pm$ 0.08 (0.11) &   43/50\\
ULIRGs		&& 1.37 $\pm$ 0.42 (1.03) &  7/50\\
(U)LIRG 	&&  0.43 $\pm$ 0.10 (0.16)  & 50\\
\hline\noalign{\smallskip}
AGN (LIRGs)		&&  0.43 $\pm$ 0.25 (0.22) &   4/11\\
AGN (ULIRGs)	&& 1.73 $\pm$ 0.67 (1.97) &  7/11\\
AGN (U)LIRG 	&&  0.91 $\pm$ 0.34 (0.29)  & 11\\
\hline\hline\noalign{\smallskip}
\end{tabular}
\vskip0.2cm\hskip0.0cm
\end{small}
%\begin{minipage}[h]{12cm}
\footnotesize
Col (1): System. Col (2): Mean (and median) total kinematic asymmetry. Col (3): Fraction of galaxies in each subsample. 
%\end{minipage}
\end{table*}

\begin{figure*}%[htbp]
\vskip-3mm
\hskip2cm\includegraphics[width=0.4\textwidth]{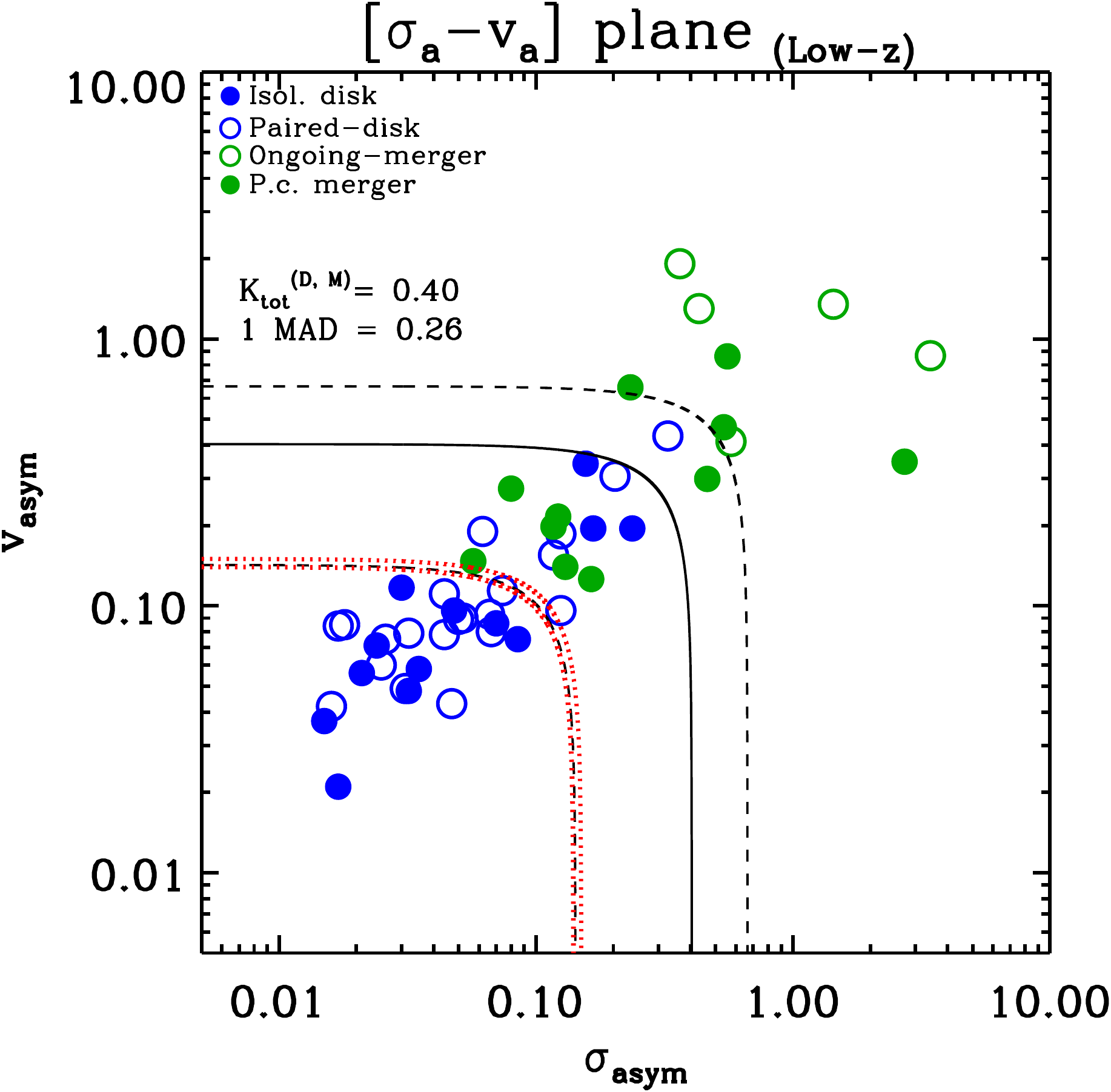}
\includegraphics[width=0.4\textwidth]{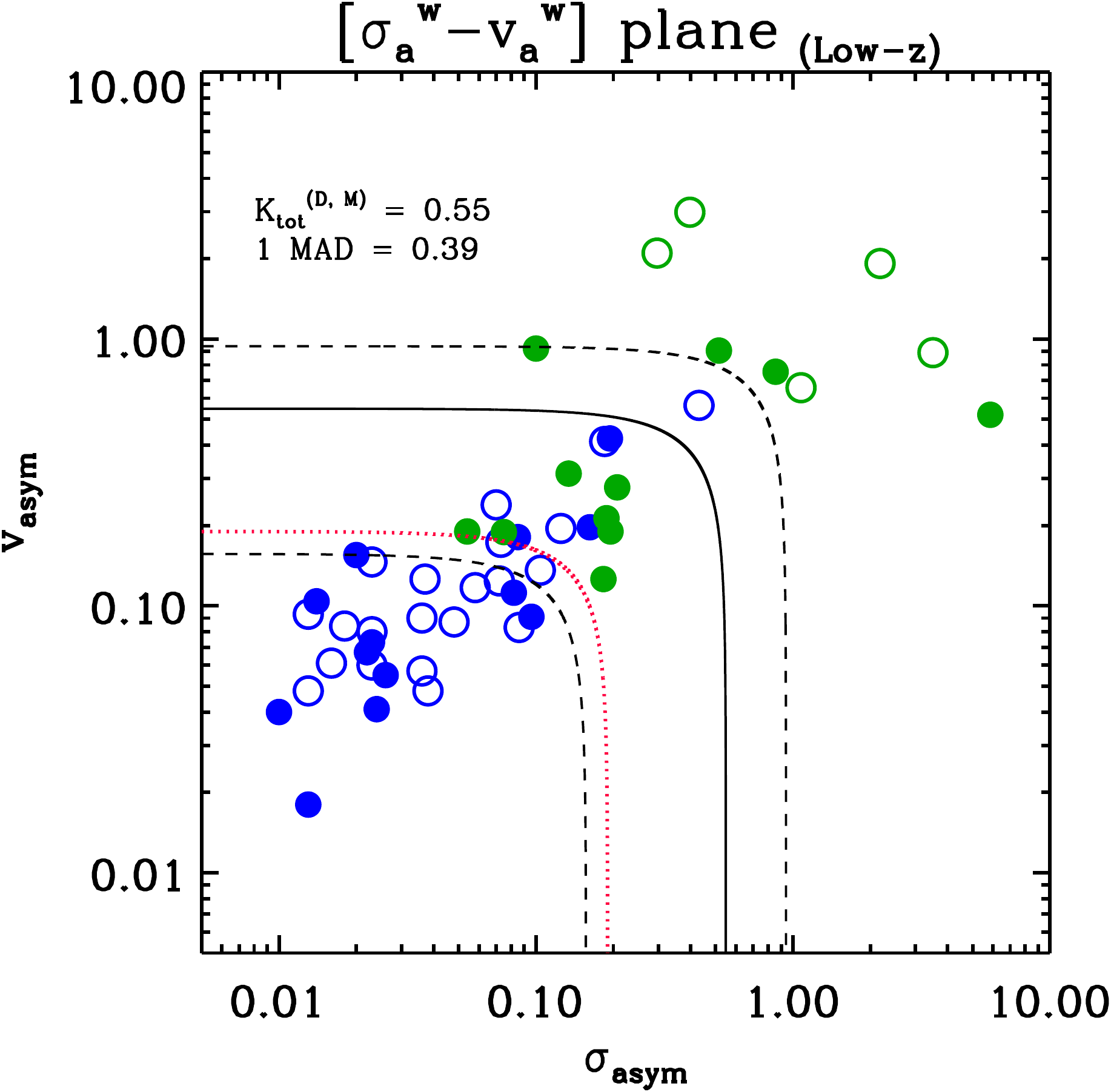}

\vskip4mm

\hskip2cm\includegraphics[width=0.4\textwidth]{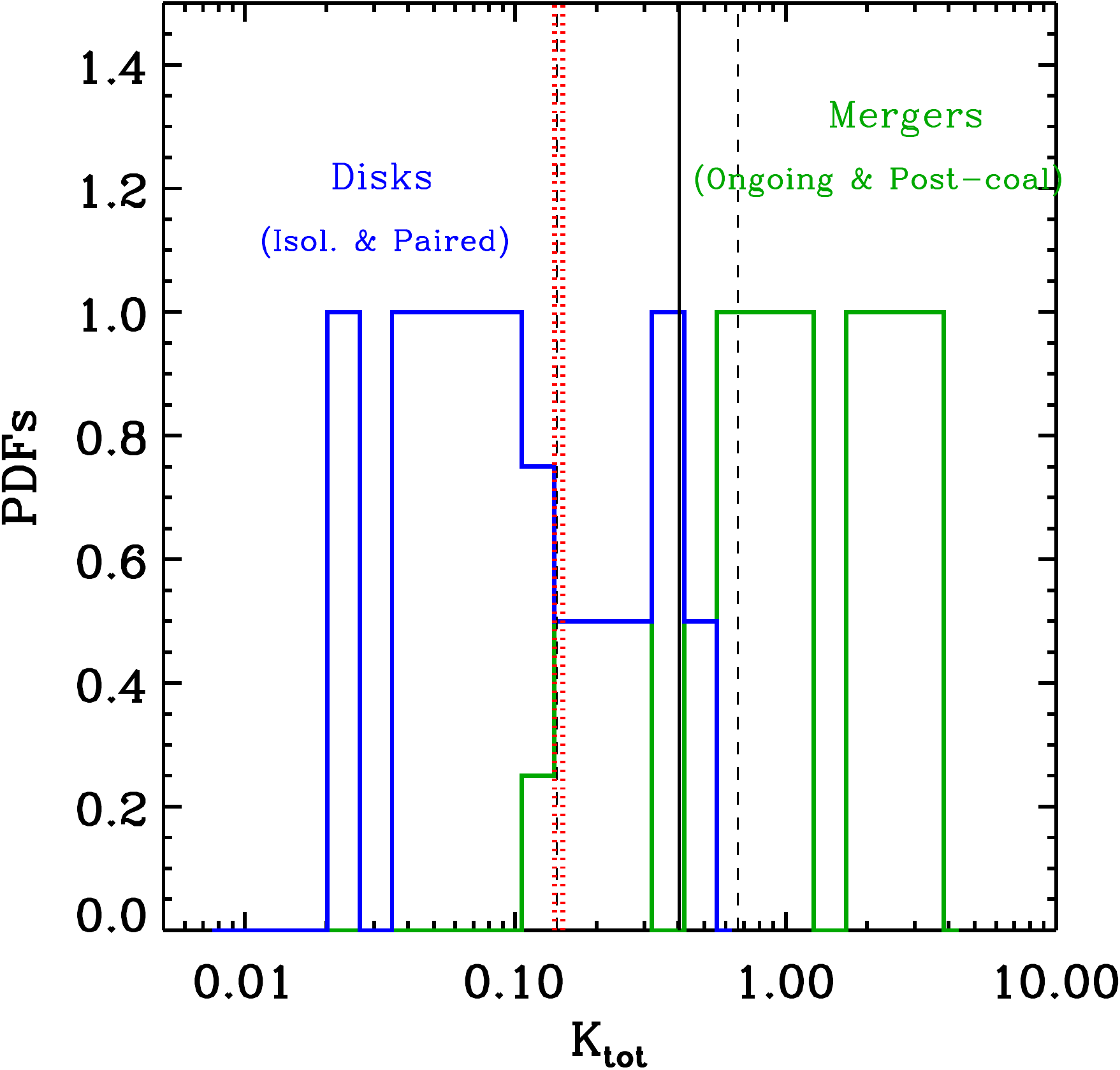}
\includegraphics[width=0.4\textwidth]{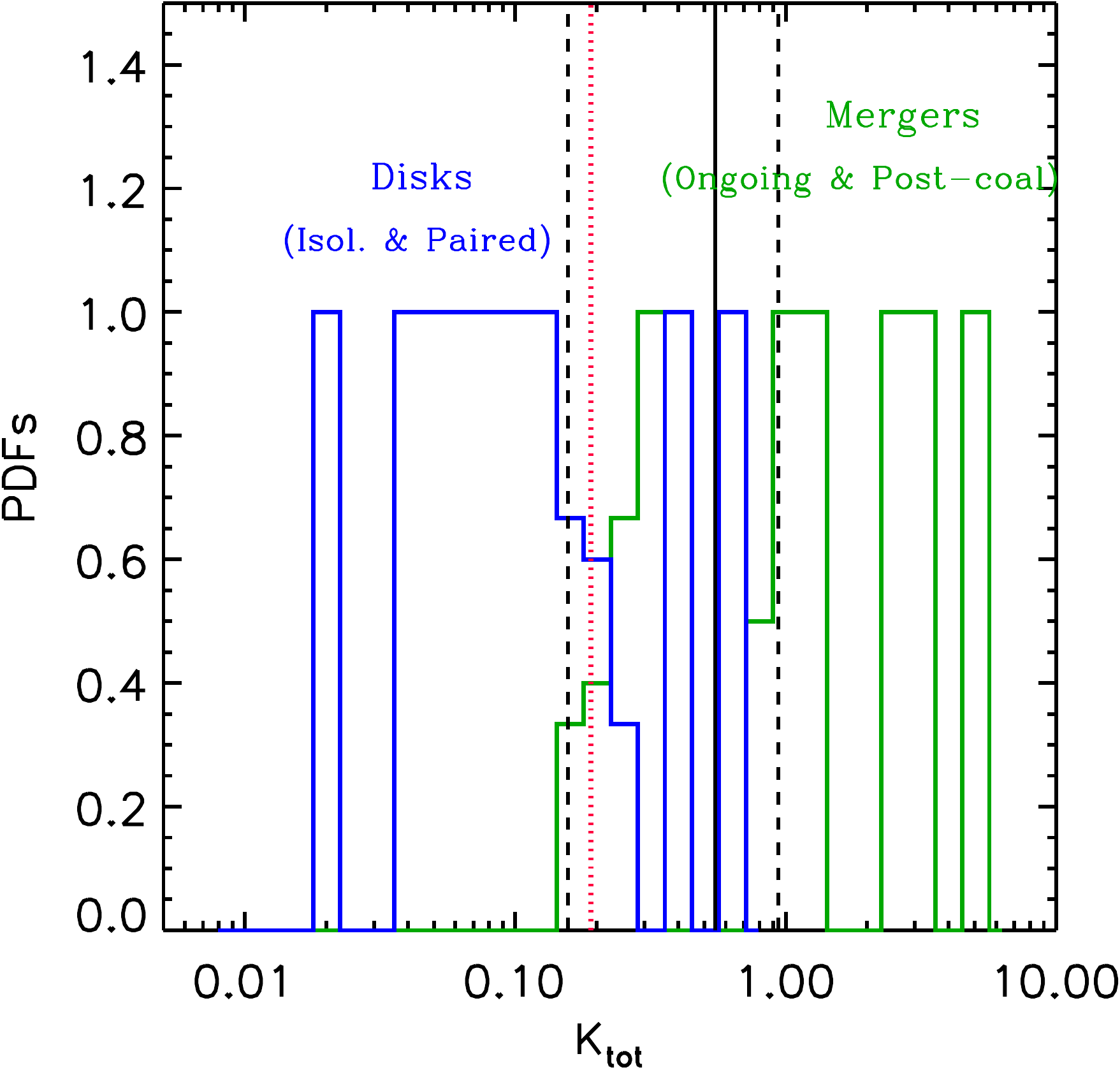}

\vskip4mm

\hskip2cm\includegraphics[width=0.4\textwidth]{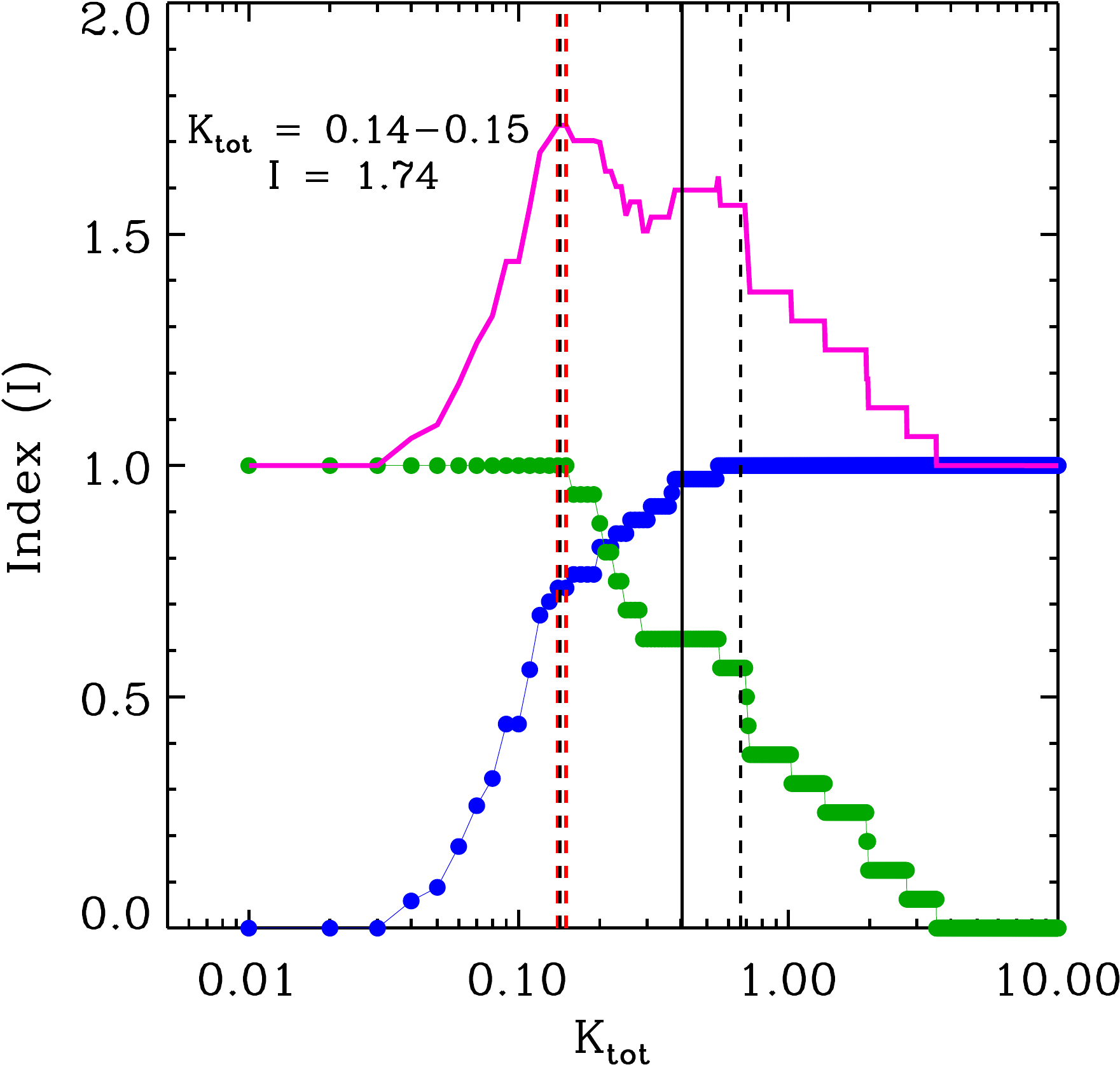}
\includegraphics[width=0.4\textwidth]{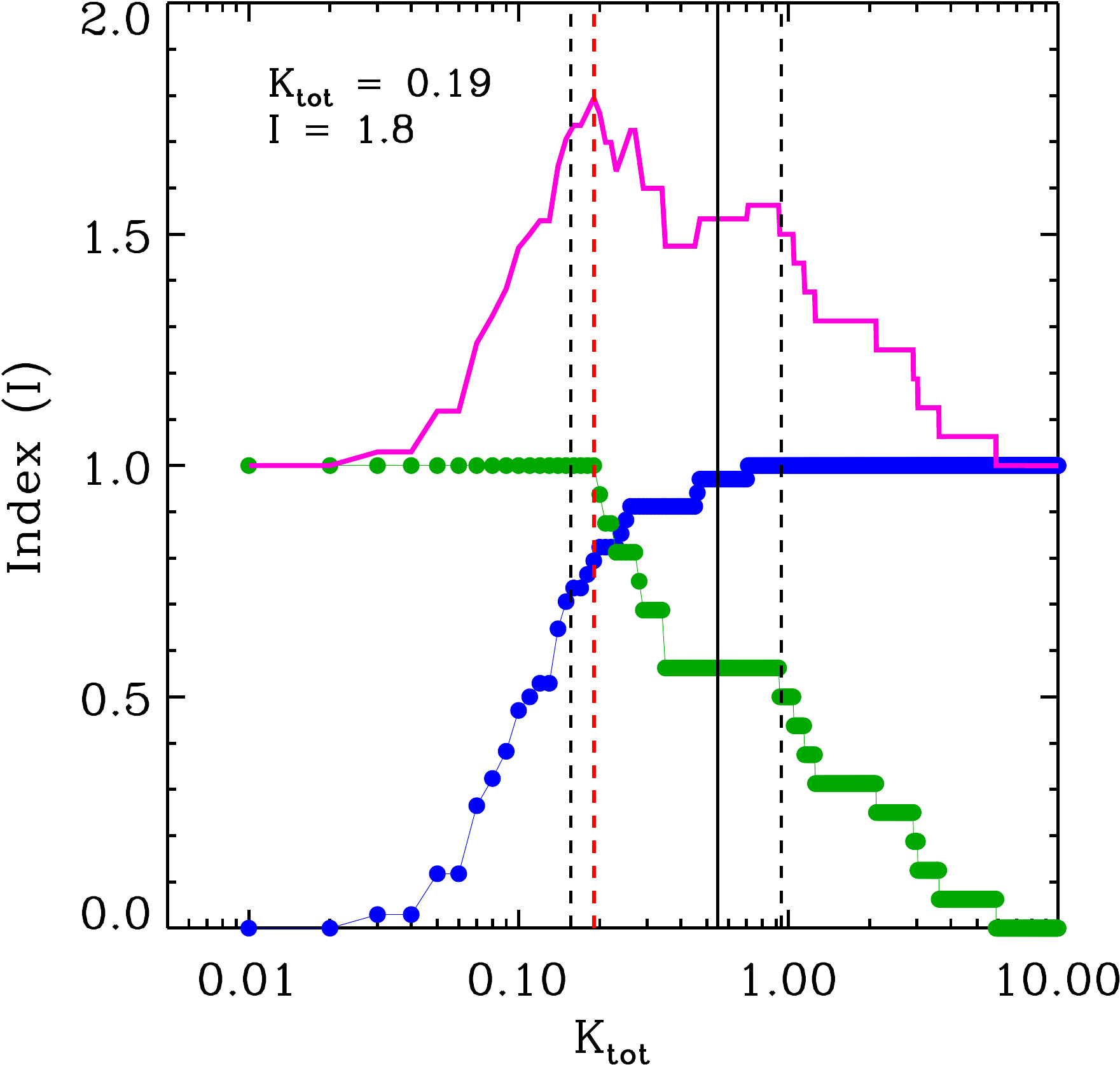}

\vskip-3mm
\hskip1cm\begin{minipage}{16.5cm}
\caption{{\bf Top:} Asymmetry measure of the velocity v$_{asym}$ and velocity dispersion $\sigma_{asym}$ fields for the whole sample observed at low--z when applying the unweighted (left) or the weighted (right) methods. Different colors distinguish the different morphological types: the solid dark blue dots are the morphologically classified isolated disks, in open blue dots the wide-pairs, in solid dark green dots the post-coalescence mergers and in open green dots the close-pairs. {\bf Middle:} The probability distribution functions (PDFs) of the respective planes normalized to the number of objects in each bin. {\bf Bottom:} Distribution of the total number of well classified galaxies at low--z as a function of the total kinematic asymmetry K$_{tot}$ in the unweighted (left) and weighted (right) planes. \newline 
In all the panels we represent the following lines: the red dashed line(s) represents the `optimal' frontier which gives us the K$_{tot}$ value for which the maximum index I is derived; in solid black line the K$_{tot}^F$ value derived according to Eq. \ref{eq:calcolo_K} while the two dashed black lines represent the statistical frontiers (K$^F_{tot}$ $\pm$ 1 MAD).}
%The K$_{tot}$ value which identifies the maximum index I is represented using a red line in all the plots. The statistical frontiers (i.e., K$^F_{tot}$ $\pm$ 1 MAD) are shown in black (dashed lines) in all the panels.
\label{MORP_all}
\end{minipage}
\end{figure*}

The kinematic asymmetry results (i.e., v$_{asym}$ and $\sigma_{asym}$) derived for the whole local sample are shown in Fig. \ref{MORP_all} when applying, respectively, the unweighted (left panels, [$\sigma_a$-v$_a$]) and weighted (right panels, [$\sigma^w_a$-v$^w_a$]) criteria. The same general trend is found in both the planes, where {\it disks} (isolated and paired disks) are characterized by lower kinematic asymmetries than {\it mergers} (ongoing and post-coalescence mergers).

In  order to distinguish {\it disks} from {\it mergers} in the two kinematic asymmetry planes we try to find out a value of the frontier applying the following approach. Since a quite large scatter is found in the asymmetry values of each kinematic class (i.e., disk and merger populations), the median value (instead of the mean) of each K$_{tot}$ distribution is considered (hereafter, K$_{tot}^{med (disk)}$ and K$_{tot}^{med (merger)}$). Then, the total kinematic asymmetry for the frontier (K$^F_{tot}$) is computed as the mean value of these two quantities:

\begin{equation}
\hspace{1cm}K^F_{tot} = \frac{1}{2} \times (K_{tot}^{med (disk)} + K_{tot}^{med (merger)}).
\label{eq:calcolo_K}
\end{equation}

The uncertainty associated to this value has been computed as the mean value of each \texttt{Median Absolute Deviation (or MAD)\footnote{The respective uncertainty associated to the $median$ value of the total kinematic asymmetry has been computed as median-absolute-deviation (hereafter, MAD). It returns a data set' s median absolute deviation from the median, that is median(|data - median(data)|). It is a proxy for the standard deviation, but is more resistant  against outliers.}} estimate associated to each distribution. As shown in Fig. \ref{MORP_all}, for K$_{tot}$ $\leq$ K$^F_{tot}$ - 1 MAD no mergers are found, while for K$_{tot}$ $\geq$ K$^F_{tot}$ + 1 MAD disks do not exist (dashed black frontiers). Indeed, the more disturbed objects are those classified as {\it ongoing-mergers} (i.e., IRAS F06035--7102, IRAS F06206--6315, IRAS F12596--1529, IRAS F22491--1808, IRAS F23128--5919)\footnote{These sources may appear as such either because they are actually in the early phase of merging or because the limited angular resolution of VIMOS does not allow to separate the contribution of each galaxy.} along with some post-coalescence sources which show high asymmetries (i.e., IRAS F05189--2524, IRAS 09022--3615, IRAS F10257--4339, IRAS F13001--2339), characterized by very disturbed kinematic maps. There is a `transition region' (i.e., K$^F_{tot}$ - 1 MAD $<$ K$_{tot}$ $<$ K$^F_{tot}$ + 1 MAD) where the distinction between {\it disk/merger} is difficult. The large dispersion and overlapping in the \texttt{kinemetry} results derived for our isolated, pre- and post-coalescence systems highlight the uncertainty in deriving a clear value of K$_{tot}$ able to clearly separate disks from mergers. The middle panels in the same figure show the corresponding probability distribution functions (PDFs) of each morphological class normalized to the number of objects in each bin.

Both the unweighted and weighted methods give similar results, although the weighted one allows to distinguish slightly better disks from mergers. This is better visible if, apart from the statistical approach, we determine the `optimal' value of the frontier able to classify our local sources in disk and merger galaxies (dashed red frontier): to this aim, the number of well classified galaxies as a function of the K$_{tot}$ is derived. Since in our sample the number of disks dominates on the number of mergers, we then define an `index' parameter (I) as the sum of the respective fractions of well classified systems in each morphological class (i.e., I$_{disk}$ for disks and I$_{merger}$ for mergers). In particular, I is defined in Eq. \ref{eq:indeces_eq}: 

$$I = I_{disk} + I_{merger} = $$

\begin{equation}
\hspace{1cm}=  \frac{\#well\hspace{1mm}class \hspace{1mm}disks}{total \hspace{1mm}\# \hspace{1mm} disks} + \frac{\# well\hspace{1mm}class\hspace{1mm}mergers}{total\hspace{1mm}\# \hspace{1mm}mergers} 
\label{eq:indeces_eq}
\end{equation}

The results are shown in the bottom panels of Fig. \ref{MORP_all} in the two cases (i.e., unweighted and weighted methods): the disk contribution is represented in blue, the merger one is shown in green and the total one in magenta. The total observed distribution allows to determine the optimal K$_{tot}$ frontier value for which the largest fraction of well classified systems (I$_{max}$) is achieved: this value is also shown in the figure. In each distribution two peaks are found and we refer to them as `main' and `secondary' peaks. The main peak I$_{max}$ identifies the K$_{tot}$ value(s) which well classifies the largest fraction of systems, where K$_{tot}(I_{max}) \sim$ 0.15 and 0.19, in the unweighted and weighted planes, respectively. As visible in the figure, the statistical K$_{tot}$ values (i.e., K$_{tot}$ $\pm$ 1 MAD) approximate well the main and secondary peaks, also defining the region where the distinction between disk/merger is difficult.

Thus, at low--z the largest value of I is derived when the weighted method is applied (i.e., I$_{max}$ = 1.8, K$_{tot}$(I$_{max}$) = 0.19), although it is only slightly larger than that derived in the unweighted plane (I$_{max}$ $\sim$ 1.7). The relative disk/merger fraction derived locally according to the weighted frontier in our sample is 27/23, implying that the number of disks almost equals that of mergers. When considering the frontier adopted by S08 (K$_{tot}$ = 0.5) the index I reaches the value of $\sim$ 1.6, clearly lower than our optimal value and close to the secondary peak. If we consider this frontier half of the post-coalescence mergers are misclassified as disks, leading to an overestimation of the disk/merger ratio. Indeed, according to their frontier the disk/merger fraction is 40/10, implying that the 80\% of our objects would be classified as {\it disks}.

However, a good agreement is found between the morphology and the kinematic classification: in particular, the I$_{disk}$ and I$_{merger}$ fractions for the low--z sample are, respectively, 79\% (74\%) vs. 100\% for the weighted (unweighted) plane when isolated and paired disks are considered as `true disks'. 

\begin{figure*}

\hskip1cm\includegraphics[width=0.4\textwidth]{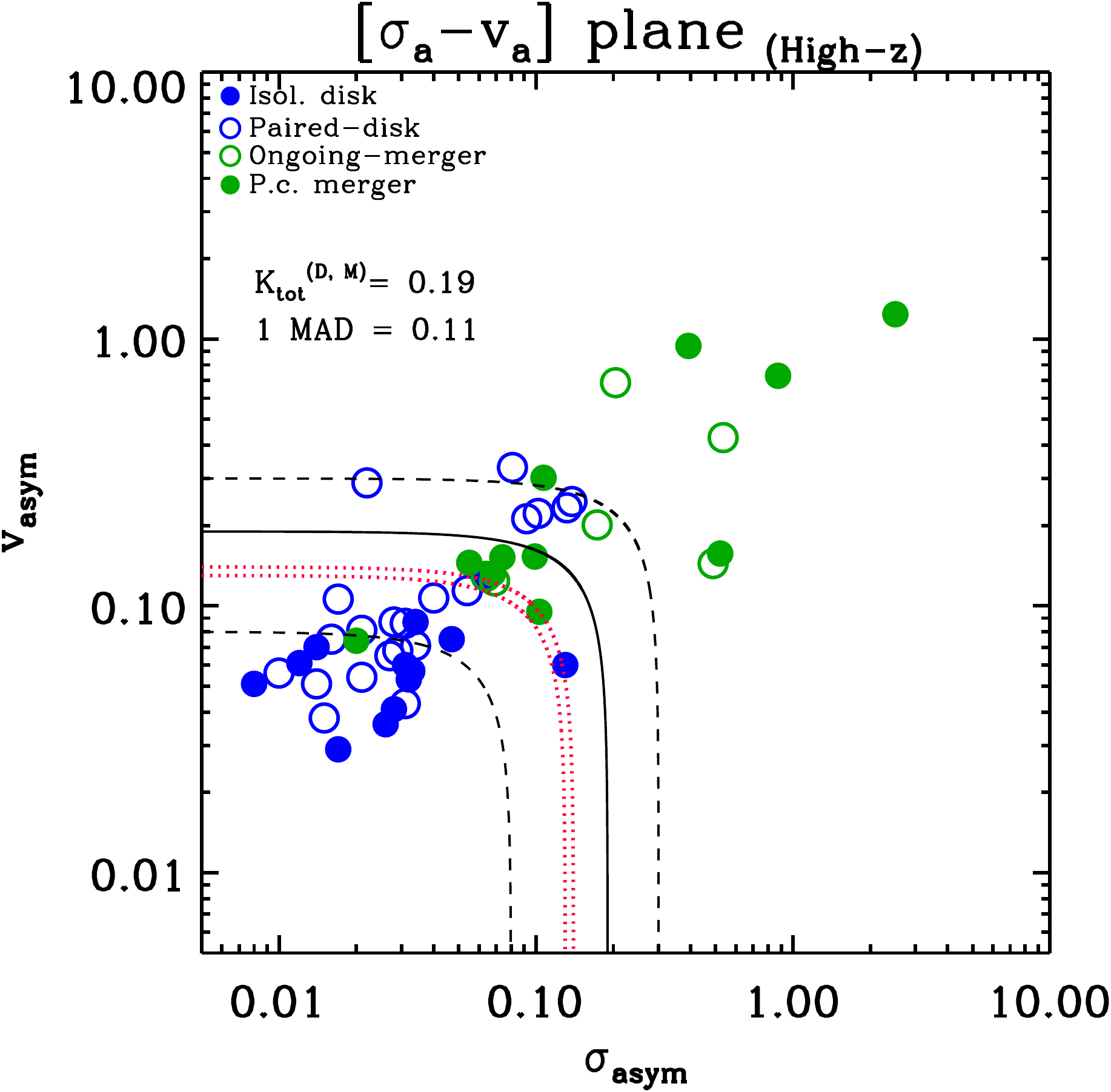}
\includegraphics[width=0.4\textwidth]{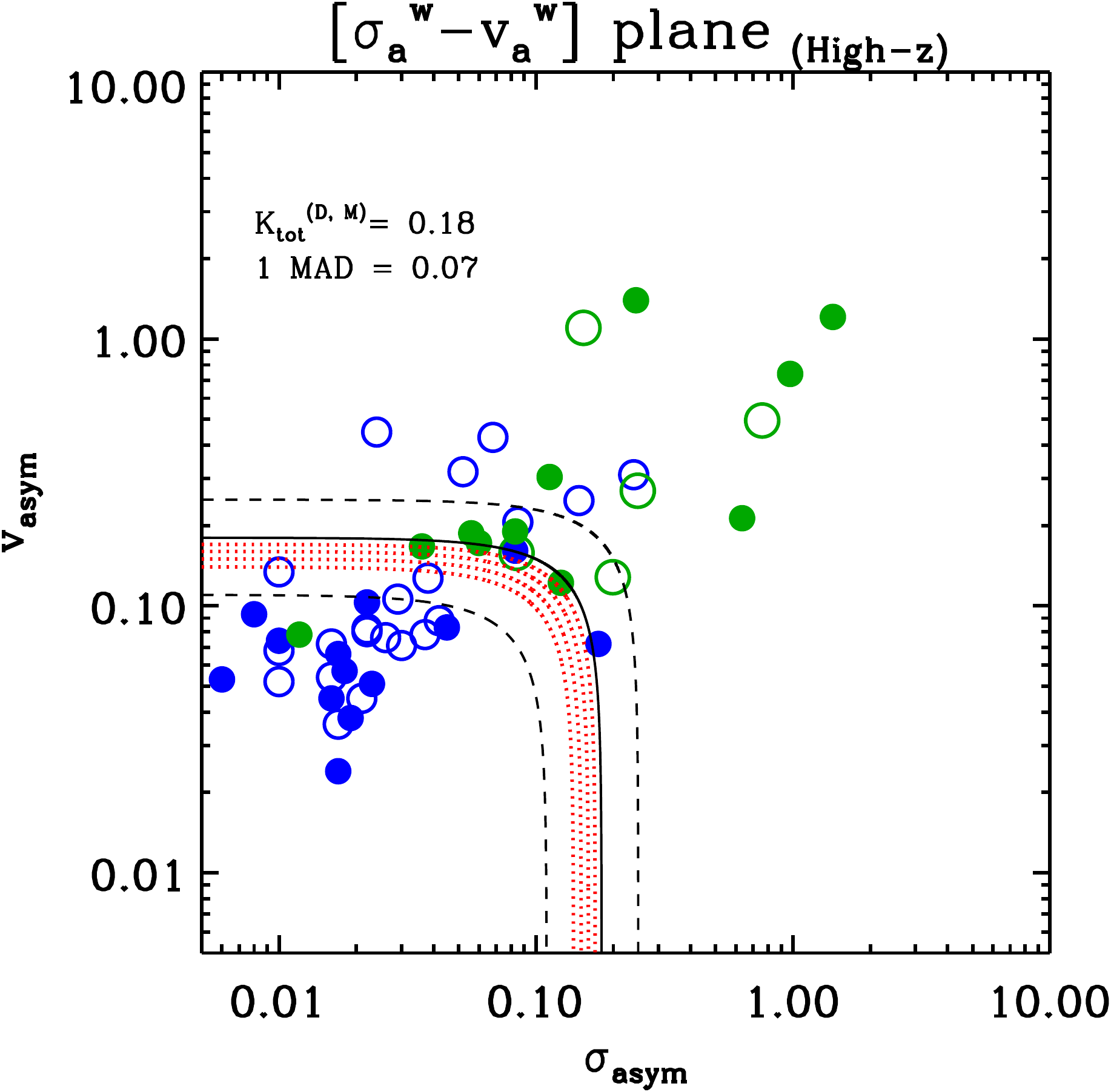}

\vskip6mm

\hskip1cm\includegraphics[width=0.4\textwidth]{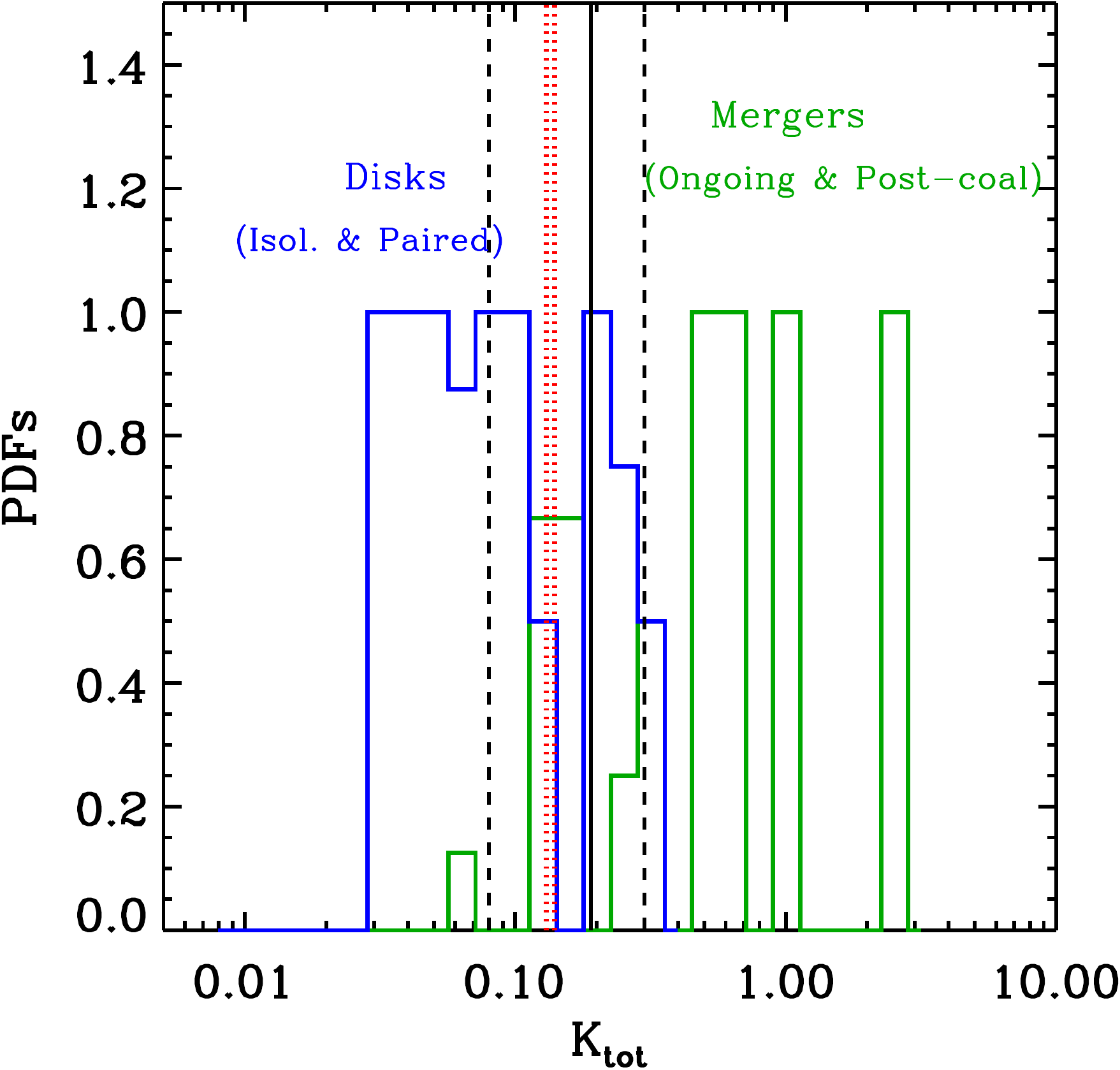}
\includegraphics[width=0.4\textwidth]{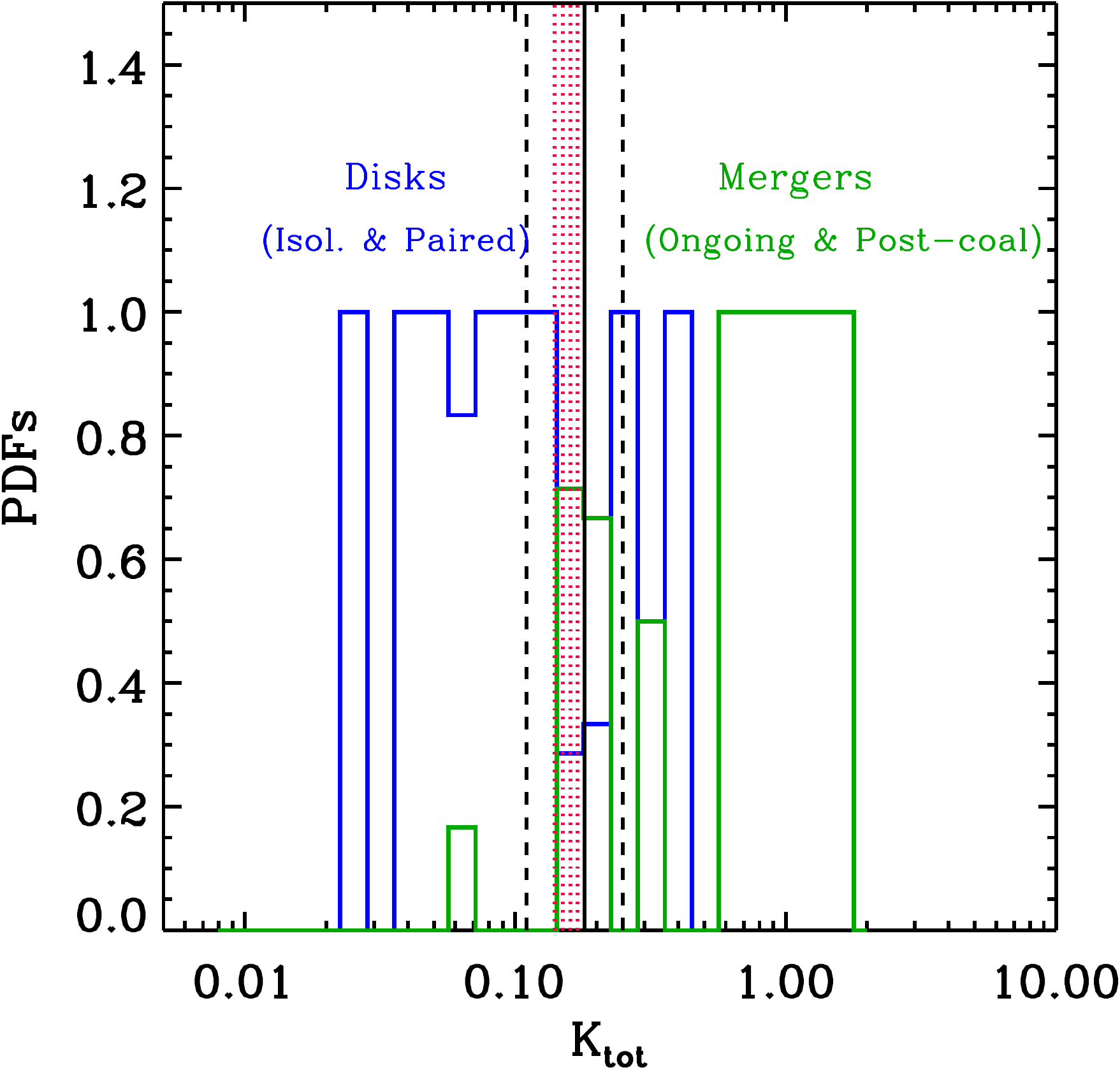}

\vskip4mm

\hskip1cm\includegraphics[width=0.4\textwidth]{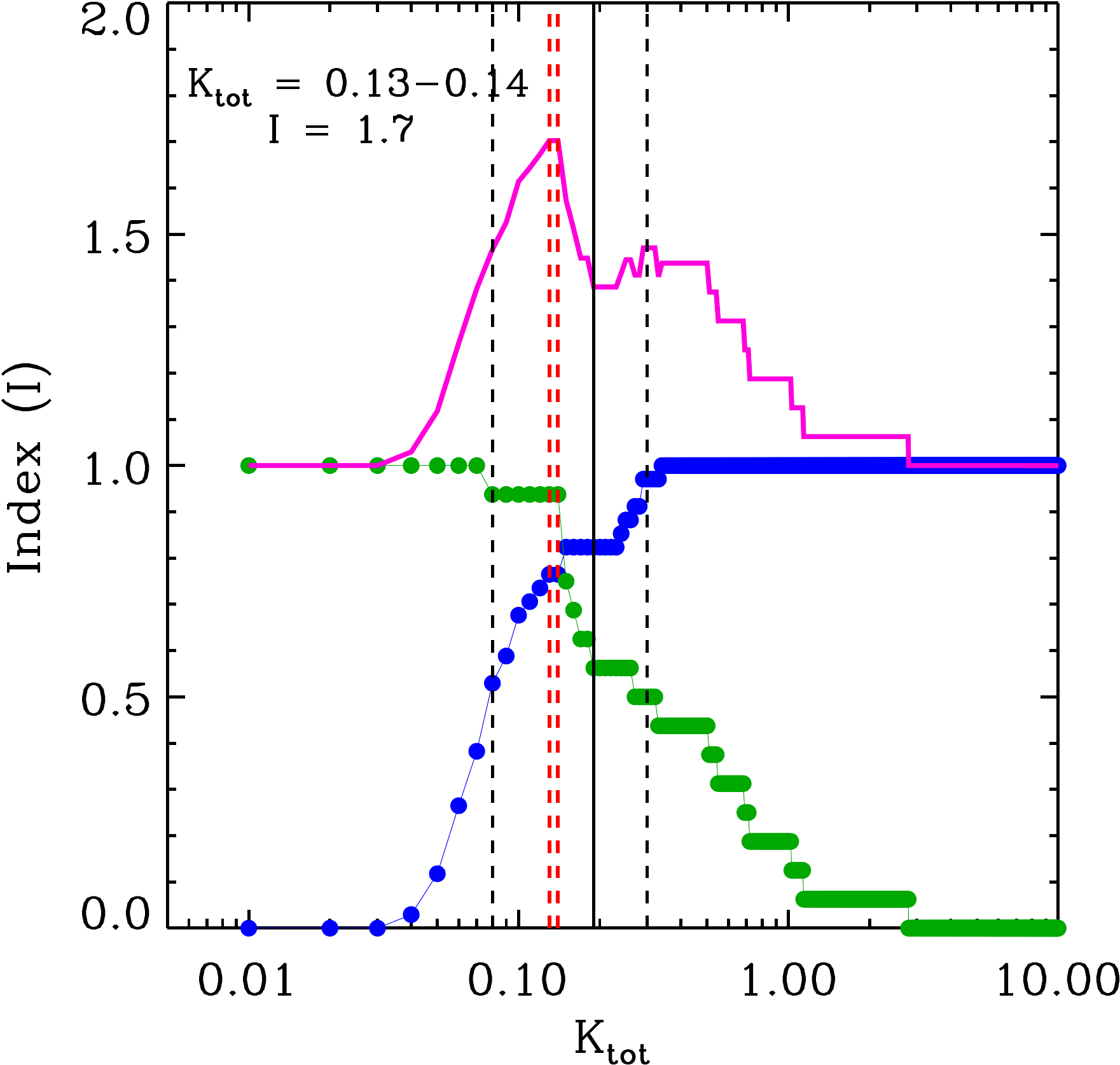}
\includegraphics[width=0.4\textwidth]{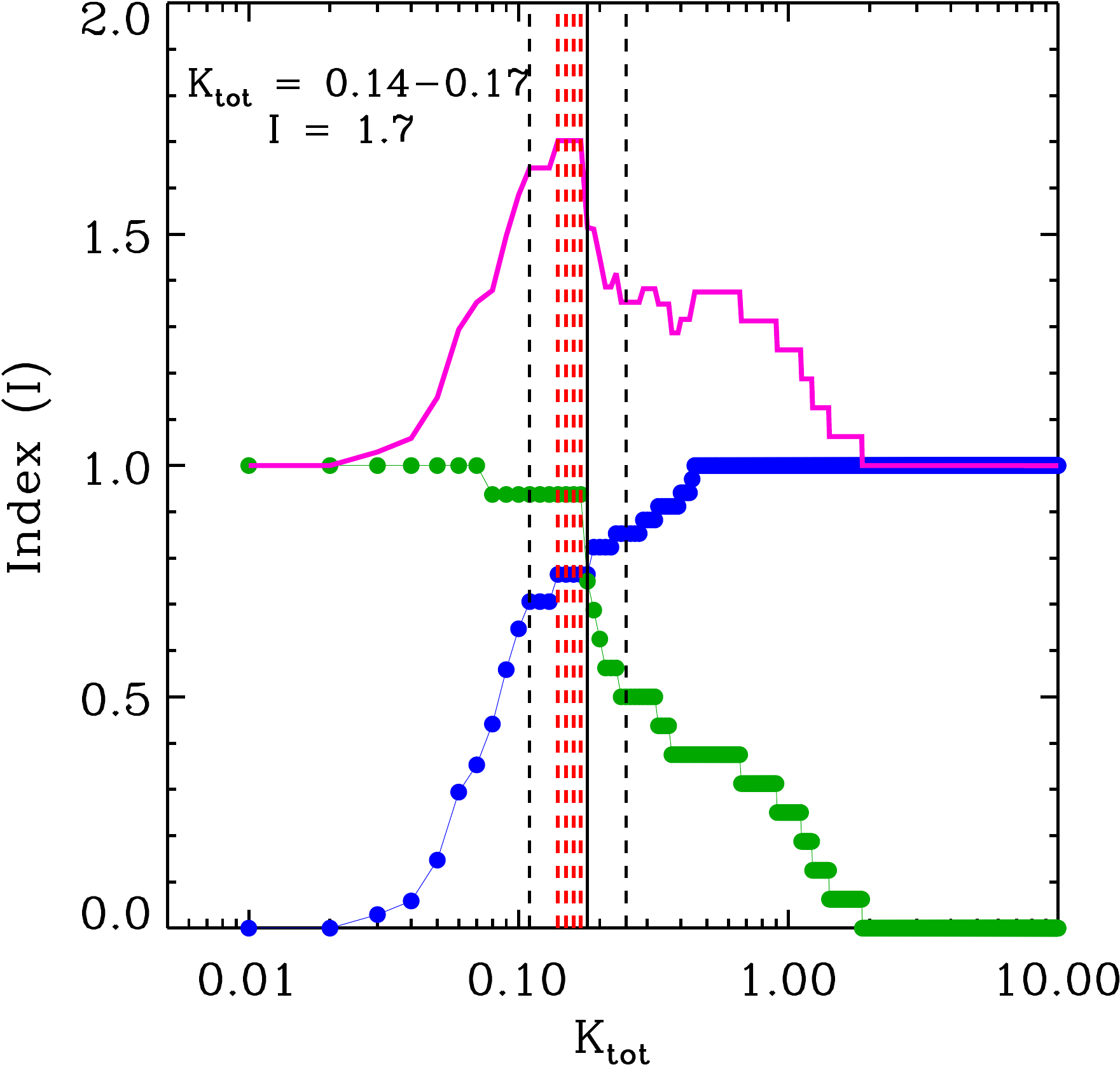}

\hskip1cm\begin{minipage}{16.5cm}
\caption{Results in the ($\sigma_a$-v$_a$) and ($\sigma_a^w$-v$_a^w$) planes for the whole sample simulated at z = 3. The panels and symbols are the same as those shown in Fig. \ref{MORP_all}.
\newline {In all the panels we represent the following lines: the red dashed lines represent the `optimal' frontiers which give us the K$_{tot}$ values for which the maximum index I is derived; in solid black line the K$_{tot}^F$ value derived according to Eq. \ref{eq:calcolo_K} while the two dashed black lines represent the statistical frontiers (K$^F_{tot}$ $\pm$ 1 MAD).}}
\label{SIM_high_morph}
\end{minipage}
\end{figure*}

\begin{figure*}
\hskip1cm\includegraphics[width=0.4\textwidth]{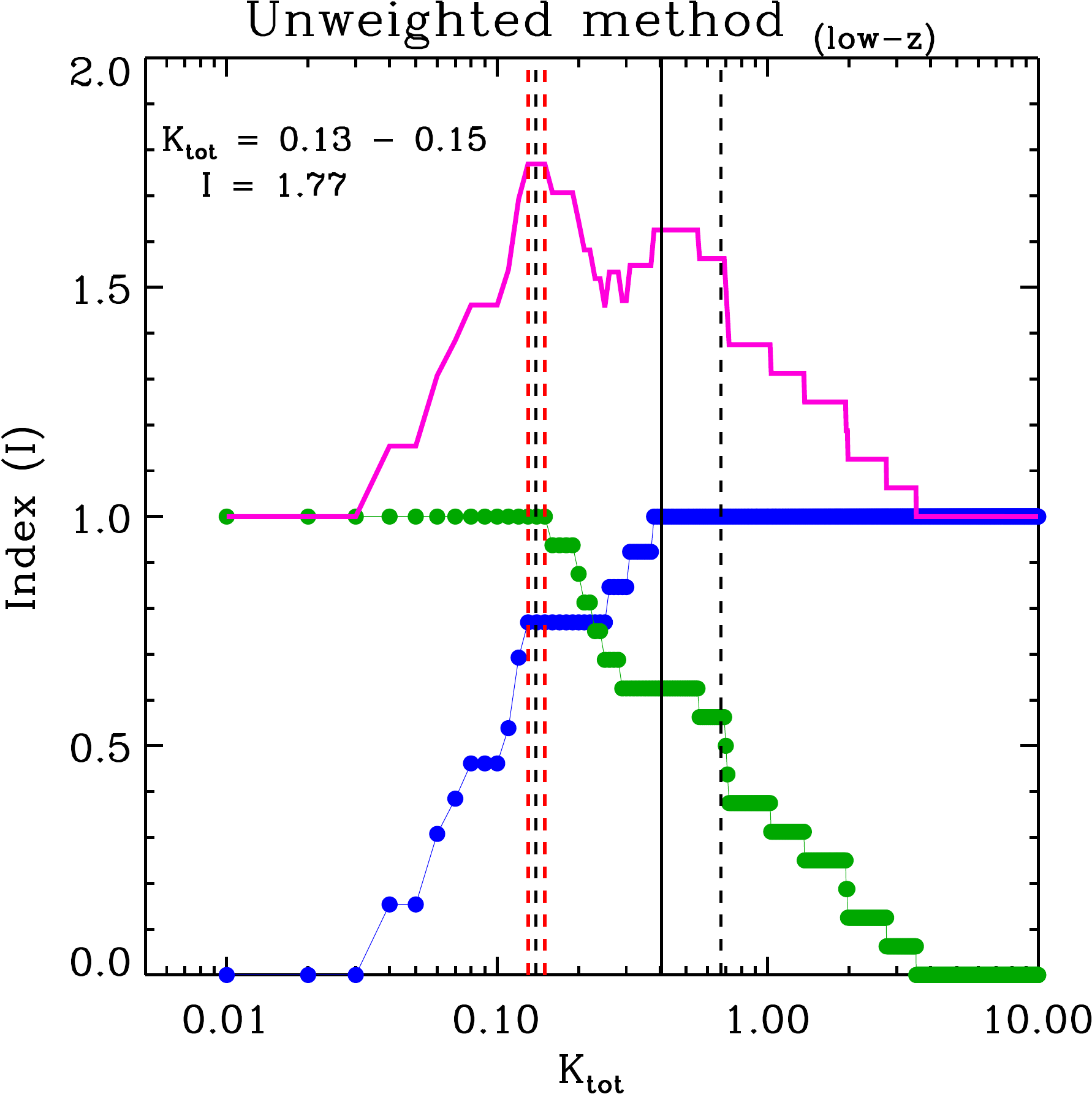}
\includegraphics[width=0.4\textwidth]{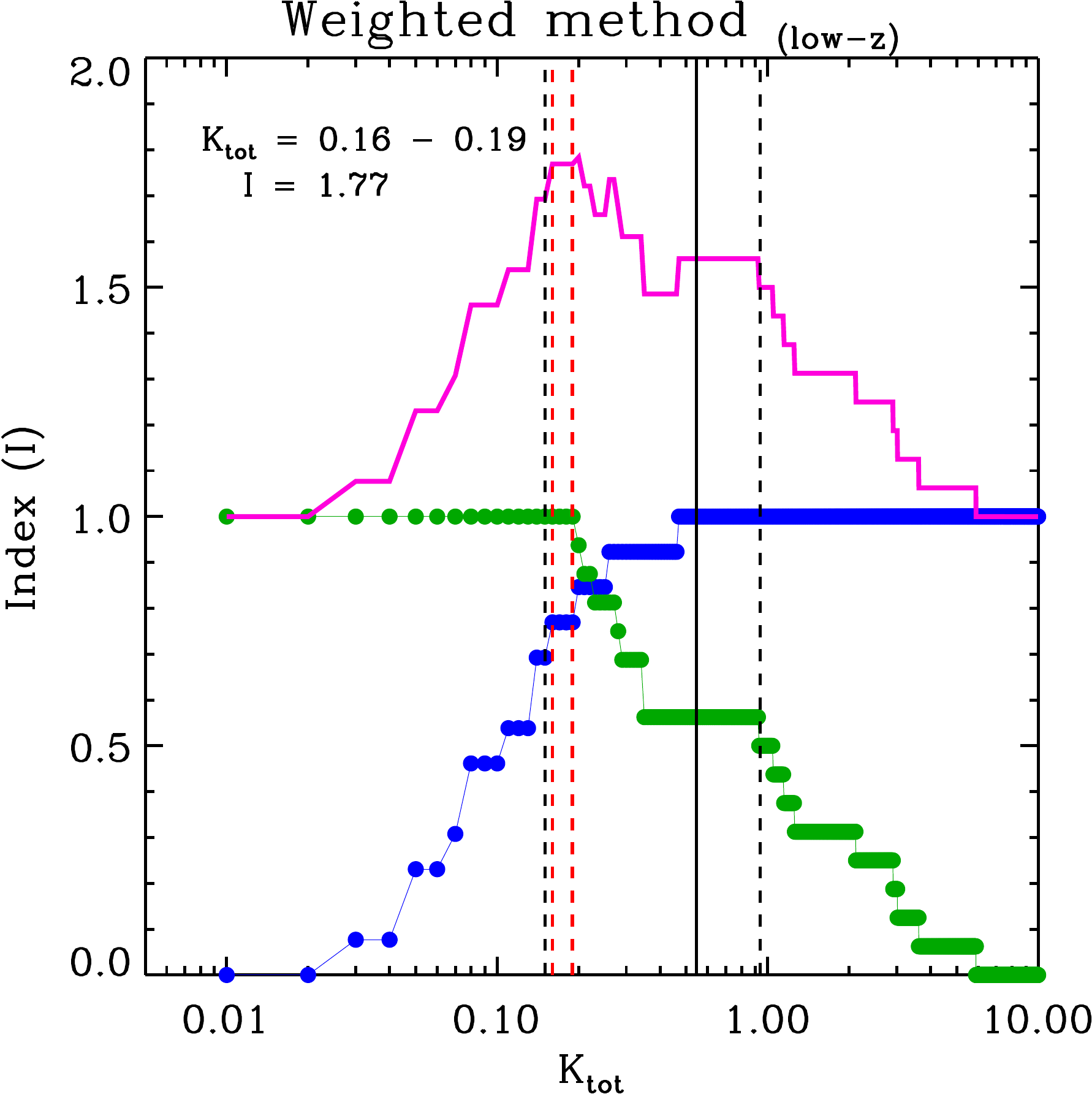}

\vskip5mm

\hskip1cm\includegraphics[width=0.4\textwidth]{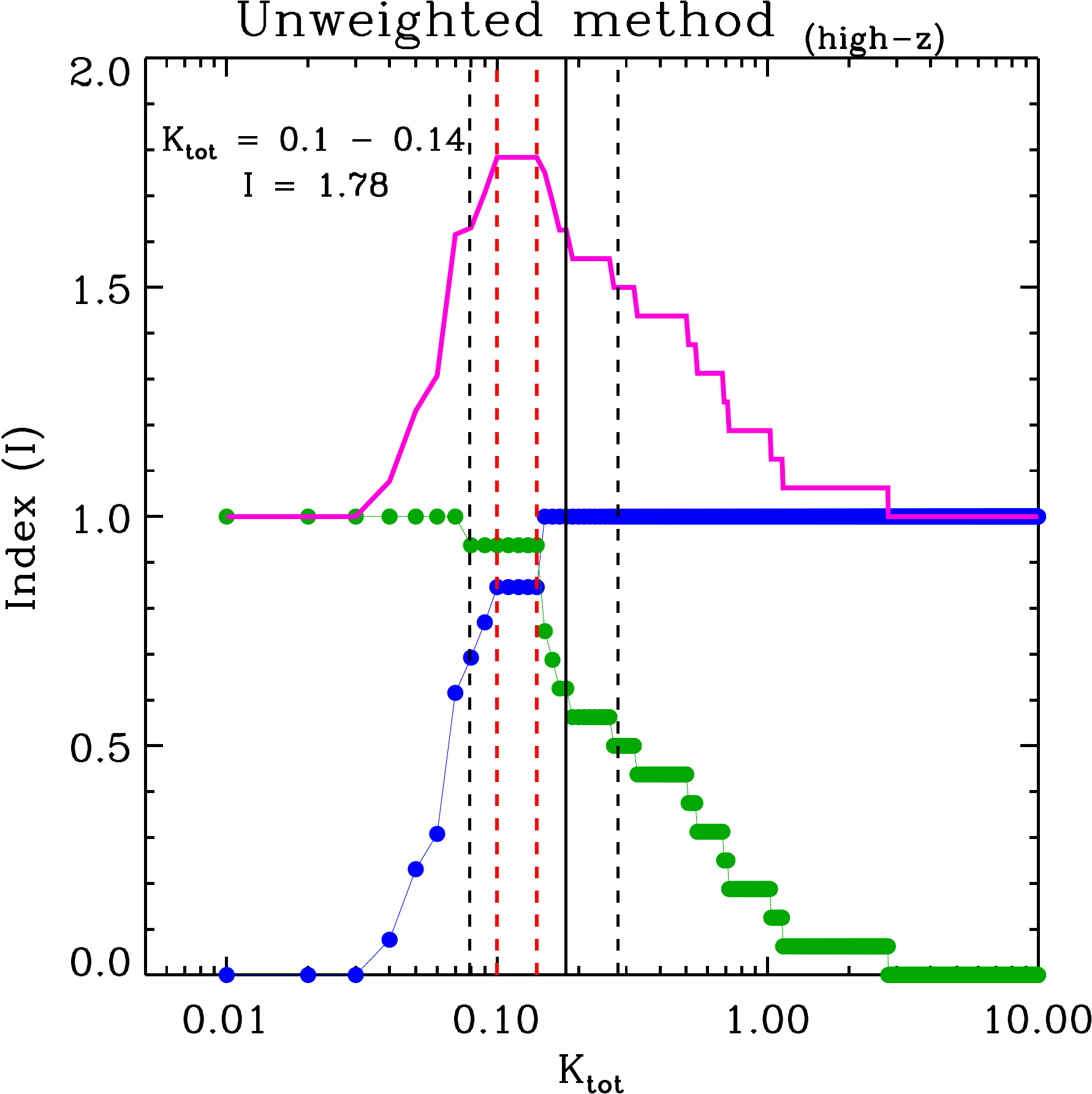}
\includegraphics[width=0.4\textwidth]{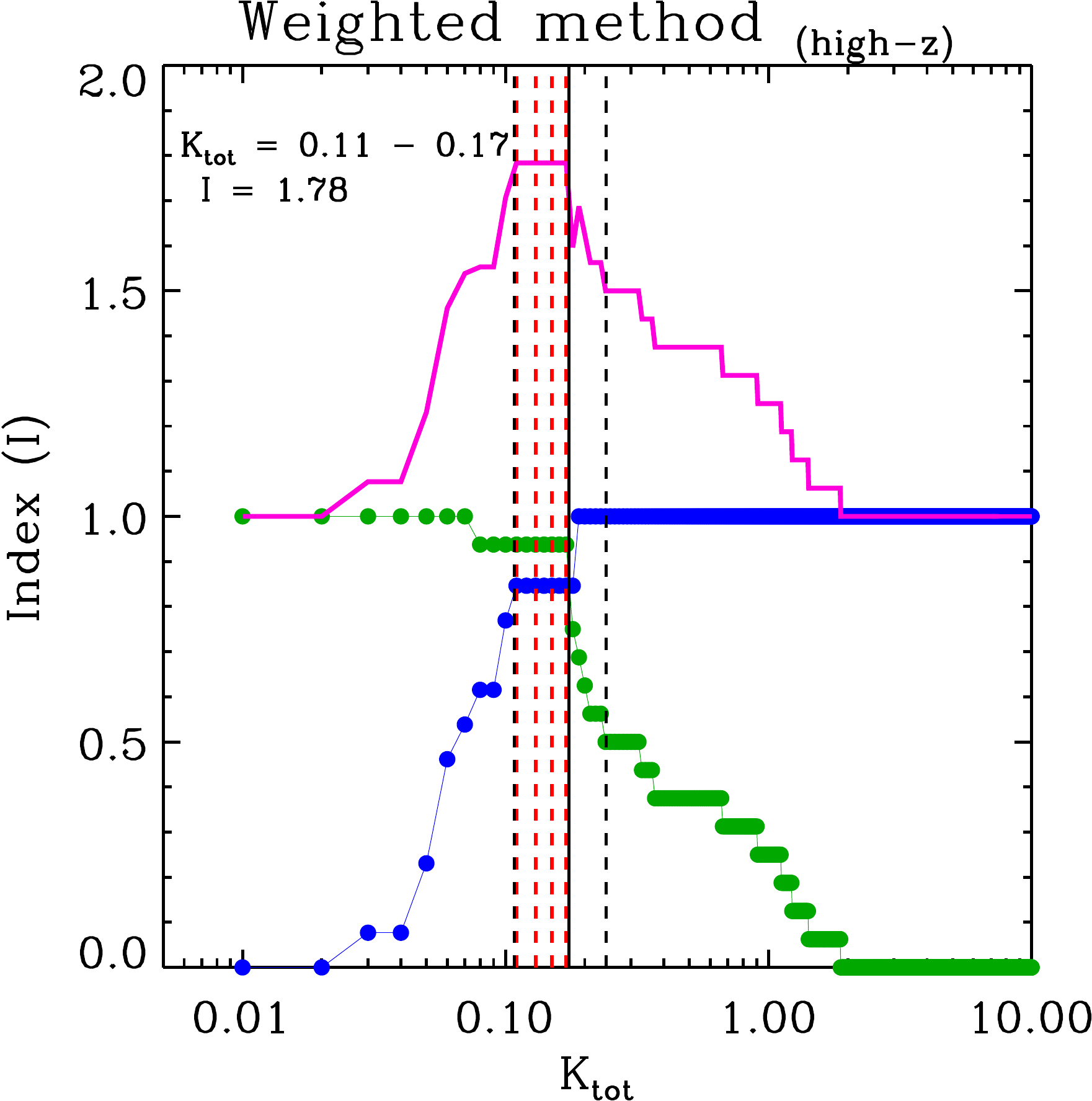}

\hskip1cm\begin{minipage}{16.5cm}
\caption{Distribution of the total number of well classified galaxies (I (magenta) = I$_{disk}$ (blue) + I$_{merger}$ (green)) at low-- (upper panels) and at high--z (bottom panels) as a function of the total kinematic asymmetry K$_{tot}$ in the unweighted (left) and weighted (right) planes when assuming that only the isolated disks are considered as `true disks' and ongoing and post--coalescence mergers are considered as `true mergers'. The red dashed lines represent the `optimal' frontier which give us the K${tot}$ values for which the maximum index I is derived; the solid black line is the K$^F_{tot}$ value derived according to Eq. 5 while the two dashed black lines represent the statistical frontiers (K$^F_{tot}$ $\pm$ 1 MAD).}
\label{indeces_mix}
\end{minipage}
\end{figure*}

If only isolated disk galaxies are considered as `true disks', the distribution of the total number of well classified galaxies (index I) as a function of K$_{tot}$ in the unweighted and weighted planes (Fig. \ref{indeces_mix}) follows the same trend than that found in the former case (bottom panels in Figs.~\ref{MORP_all}, \ref{SIM_high_morph}).

In Tab. \ref{comparison_tabl_} a comparison of the different (morphological and kinematic) classifications of the (U)LIRG sample is summarized.

\begin{table*}
\centering
\caption{Comparison of the different (morphological and kinematic) classifications of the local (U)LIRG sample.}
\label{comparison_tabl_}
\begin{scriptsize}
\begin{tabular}{l c c c l} \\
\hline\hline\noalign{\smallskip}  
ID1{} &  Morphological   &  Visual kinematic    &  \texttt{Kinemetry}   &  \texttt{Kinemetry}    \\
IRAS &  classification &   classification  &  value of K$_{tot}$ &  classification  \\ 
(1) & (2) & (3) & (4)  & (5) \\
\hline\noalign{\smallskip} 	
F01159-4443 S    & 1 & PD  &  0.25  & \texttt{Disk$^\star$}\\ 
F01159-4443 N   	&  1   & PD & 0.12 & \texttt{Disk} \\ 
\hline\noalign{\smallskip} 	
F01341-3735 S   	& 1   & PD & 0.14 & \texttt{Disk}\\
F01341-3735 N   	& 1   & PD & 0.06 & \texttt{Disk}\\
\hline\noalign{\smallskip} 	
F04315-0840    & 2 	& CK &  0.2 & \texttt{Disk$^\star$}\\
\hline\noalign{\smallskip} 	
F05189-2524     	& 2 	& CK &  1.14 & \texttt{Merger}\\
\hline\noalign{\smallskip} 	
F06035-7102    		& 1 (ongoing) 	& CK & 2.91 & \texttt{Merger} \\
\hline\noalign{\smallskip} 	
F06076-2139  S   	& 1 	& PD & 0.1 & \texttt{Disk}\\
F06076-2139  N   	& 1 	& PD & 0.09 & \texttt{Disk}\\
\hline\noalign{\smallskip} 	
F06206-6315    		& 1 (ongoing) 	& PD & 1.26 &\texttt{Merger}\\
\hline\noalign{\smallskip} 	
F06259-4780 S  & 1 	& RD & 0.09 &\texttt{Disk}\\
F06259-4780 C  & 1 	& RD & 0.10 &\texttt{Disk}\\
F06259-4780 N  & 1 	& RD & 0.13 &\texttt{Disk}\\
\hline\noalign{\smallskip} 	
F06295-1735   	& 0  & PD & 0.20 & \texttt{Disk$^\star$}\\
\hline\noalign{\smallskip} 	
F06592-6313    	& 0 	& PD &  0.13 & \texttt{Disk}\\
\hline\noalign{\smallskip} 	
F07027-6011 S   & 0 & RD & 0.14 & \texttt{Disk}\\
F07027-6011 N   & 0 & PD & 0.06 & \texttt{Disk}\\
\hline\noalign{\smallskip} 	
F07160-6215  & 0 	& PD (CK)& 0.47 & \texttt{Disk$^\star$}\\
\hline\noalign{\smallskip} 	
08355-4944     	& 2 	& PD & 0.27 & \texttt{Disk$^\star$}\\
\hline\noalign{\smallskip} 	
08424-3130 S   & 1 	& PD & 0.17 & \texttt{Disk}\\
08424-3130 N   & 1 	& --  & --& -- \\
\hline\noalign{\smallskip} 	
F08520-6850 E  & 1 	& RD &  0.45 & \texttt{Disk$^\star$}\\
F08520-6850 W & 1 	& PD (RD) & 0.07 & \texttt{Disk}\\ 
\hline\noalign{\smallskip} 	
09022-3615    	& 2 	& CK & 1.04 & \texttt{Merger}\\
\hline\noalign{\smallskip} 	
F09437+0317 S  & 1 (0) & RD &  0.08 &\texttt{Disk}\\
F09437+0317N  &  1 (0) & RD & 0.06 & \texttt{Disk}\\
\hline\noalign{\smallskip} 	
F10015-0614 & 0 	& PD &  0.10 & \texttt{Disk} \\
\hline\noalign{\smallskip} 	
F10038-3338   	& 2 	& CK &  0.35 & \texttt{Disk$^\star$} \\ 
\hline\noalign{\smallskip} 	
F10257-4339  	& 2 	& PD & 5.88 & \texttt{Merger}\\
\hline\noalign{\smallskip} 	
F10409-4556  		& 0 	& RD & 0.16 & \texttt{Disk} \\
\hline\noalign{\smallskip} 	
F10567-4310  		& 0 	& RD & 0.05 & \texttt{Disk} \\
\hline\noalign{\smallskip} 	
F11255-4120   	& 0 	& PD &   0.07 & \texttt{Disk} \\
\hline\noalign{\smallskip} 	
F11506-3851  	& 0 	& RD &   0.08 & \texttt{Disk}\\
\hline\noalign{\smallskip} 	
F12043-3140 S  	& 1	& PD & 0.13 & \texttt{Disk}\\ 
F12043-3140 N 	& 1	& PD (CK) & 0.71 & \texttt{Merger$^\star$} \\ 
\hline\noalign{\smallskip} 	
F12115-4656 & 0	& RD & 0.02 & \texttt{Disk}  \\ 
\hline\noalign{\smallskip} 	
12116-5615 	& 2 (0)& PD &   0.28 & \texttt{Disk$^\star$}\\
\hline\noalign{\smallskip} 	
F12596-1529  	& 1 (ongoing) 	& - &  3.01 & \texttt{Merger}\\
\hline\noalign{\smallskip} 	
F13001-2339 & 2 (0/1) & CK & 0.93 & \texttt{Merger$^\star$} \\
\hline\noalign{\smallskip} 	
F13229-2934  & 0 	& CK &   0.25 & \texttt{Disk$^\star$} \\
\hline\noalign{\smallskip} 	
F14544-4255 E  	& 1 & PD &   0.15 & \texttt{Disk}\\
F14544-4255 W 	& 1 & CK (PD) & 0.23 & \texttt{Disk$^\star$}  \\
\hline\noalign{\smallskip} 	
F17138-1017 	& 2 (0)& PD &  0.22 &\texttt{Disk$^\star$} \\
\hline\noalign{\smallskip} 	
F18093-5744 S  	& 1 &  RD &   0.06 & \texttt{Disk}	\\
F18093-5744 C  	& 1 &  CK (PD) &  0.19 & \texttt{Disk$^\star$}	\\
F18093-5744 N  	& 1 &  RD &    0.05 & \texttt{Disk}   \\
\hline\noalign{\smallskip} 	
F21130-4446  & 2 	& CK &	0.34 & \texttt{Disk$^\star$}	\\
\hline\noalign{\smallskip} 	
F21453-3511   	& 2 	& PD &  0.20 & \texttt{Disk$^\star$}  \\
\hline\noalign{\smallskip} 	
F22132-3705  	& 0 	& RD & 0.04 & \texttt{Disk}  \\
\hline\noalign{\smallskip} 	
F22491-1808  	& 1 (ongoing)  &  CK (PD) &	0.95	 & \texttt{Merger}\\
\hline\noalign{\smallskip} 	
F23128-5919  	& 1 (ongoing)  &  CK &  3.62 & \texttt{Merger}\\
\hline\hline
\end{tabular}
\end{scriptsize}
\begin{minipage}{18cm}
   \vskip0.01cm\hskip0.0cm
\footnotesize
\tablefoot
{Col (1): Object designation in the Infrared Astronomical Satellite (IRAS) Faint Source Catalog (FSC). 
Col (2): Morphological class as defined in previous works (i.e., A08, MI10, RZ11, B13) as follows: 0 identifies {\it isolated} object, 1 {\it pre-coalescence} system, and 2 stands for post-coalescence {\it merger}. 
Col (3): Visual kinematic classification as in B13. \texttt{RD} stands for \texttt{rotating disk}, \texttt{PD} \texttt{perturbed disk} and \texttt{CK} are systems with \texttt{complex kinematics}.
Col (4): \texttt{Kinemetry} value of the total kinematic asymmetry K$_{tot}$ defined in Sect. 3.4 for the observed systems as derived according to the `weighted' method. 
Col (5): \texttt{Kinemetry} classification of the observed systems according to the `weighted' method (K$_F$ = 0.19). Galaxies classified as \texttt{disk$^\star$} or \texttt{merger$^\star$} are those lying in the transition region (see Sect. 3.4). No \texttt{kinemetry} analysis has been performed for the galaxy IRAS 08424-3130 N since it is located in the edge of the VIMOS FoV.}
\end{minipage}
\end{table*}

\subsection{Total kinematic asymmetries of high--z simulated (U)LIRGs}
\label{blabla_high}

We apply \texttt{kinemetry} to the high--z simulated kinematic maps (see Sect.\ref{SIMS}) and the results are shown in Fig.~\ref{SIM_high_morph}, respectively, for the unweighted (left panels, [$\sigma_a$-v$_a$]$_{High-z}$) and weighted (right panels, [$\sigma^w_a$-v$^w_a$]$_{High-z}$) planes. As expected, the results are characterized by lower kinematic asymmetries than those obtained locally (e.g., \citealt{Gon10}, B12). This result is due to the fact that when lowering the linear resolution the kinematic deviations are smoothed, making objects to appear more symmetric than they actually are. However, a few sources among the ongoing- (i.e., IRAS F06035--7102, IRAS F2249--1808, IRAS F23128--5919) and post-coalescence (i.e., IRAS F09022--3615, IRAS F05189--2524, IRAS F10257--4339, IRAS F13001--2339) mergers still preserve quite high values.

Thus, the distribution of the number of well classified objects as a function of the K$_{tot}$ is considered (bottom panels in Fig. \ref{SIM_high_morph}). The main and secondary peaks are identified as well. The maximum index I of 1.7 is reached in both the planes where the K$_{tot}$ assumes the (average) value of $\sim$ 0.14 and $\sim$ 0.16 in the unweighted and weighted planes, respectively. We derive a disk/merger ratio at high--z of 26/24, which is approximately the same ratio found locally. The value adopted by S08 would imply an index I of $\sim$ 1.4, clearly lower than the optimal value derived by us. In this case about two thirds of our `mergers' would be classified as `disks', largely overestimating the disk/merger ratio in our sample. In such a case the total number of disks would largely exceed the number of mergers by a factor of 7 (44 disks -- 6 mergers, 12\% mergers).

For the high--z simulated sample the derived I$_{disk}$ and I$_{merger}$ fractions result, respectively, in 76\% and 94\% (for both the methods) when isolated and paired disks are considered as `true disks'. If we exclude the paired disks from the `true disk' group, similar results are derived as well (i.e., 85\% and 94\%; Tab. \ref{well_corr_class}).

\subsection{Comparison between low-- and high--z \texttt{kinemetry} results in our (U)LIRG sample}
\label{aggiunto}

As a result, the comparison between local and high--z results obtained using both the unweighted and weighted methods allows to draw the following conclusions:

\begin{itemize}

\item At low--z similar results are found for the unweighted and the weighted methods, although for the weighted one `disks' are slightly better separated from `mergers'. The optimal K$_{tot}$ value able to classify the largest number of objects is $\sim$ 0.19:  according to this result the derived disk/merger fraction found locally is 27/23 (54\% disks, 46\% mergers).

\vskip2mm

\item A `transition region' ($|$ K$_{tot}$ - K$_{tot}^F$ $|$ $<$ 1 MAD) where the disk/merger classification in uncertain is found in the asymmetry plane with the total kinematic values (at low--z) in the range 0.16 (0.14) $<$ K$_{tot}$ $<$ 0.94 (0.66) for the weighted (unweighted) plane(s). Outside this range ($|$ K$_{tot}$ - K$_{tot}^F$ $|$ $\geq$ 1 MAD) we are able to well classify {\it disks} and {\it mergers}. 

\vskip2mm

\item At high--z, a trend similar to that found locally is obtained but characterized by lower total kinematic asymmetries K$_{tot}$ as a consequence of the resolution effects. Slightly better results are derived when using the weighted method, in which the main peak in the I distribution is better defined, with a K$_{tot}$ $\sim$ 0.16 and a resulting disk/merger ratio of 26/24 (52\% disks, 48\% mergers); 

\vskip2mm

\item If the frontier obtained by S08 (K$_{tot}$ = 0.5) is considered, the fraction of well classified objects (I) would be clearly lower down to 1.6 -- 1.4 with respect to our optimal values (I $\sim$ 1.8 -- 1.7), respectively, at low-- and at high--z. The S08 limit implies that at least half of the post-coalescence mergers would be misclassified as disks thus leading to an overestimation of the disk/merger ratio, with more than 80\% of the sources classified as `disk'.

\vskip2mm

\item  
If only isolated disks are considered as `true disks', the index distribution I as a function of K$_{tot}$ in both the planes, at low-- and at high--z, follows the same general trend than that derived when isolated and paired disks are considered as `true disks'. In both the cases, the derived fractions of `well classified disks' vs. `well classified mergers' according to the two methods give similar results both locally ($\sim$80\% vs. 100\%) and at high--z (75--85\% vs. 94\%; Tab. \ref{well_corr_class}). 

\end{itemize}

\begin{table*}
\caption {Comparison of the values of correctly identified disks (I$_{disk}$) and mergers (I$_{merger}$) along with their associated K$_{tot}$ (optimal) values when using the `unweighted' and `weighted' methods at low-- and at high--z. All these values are computed in the cases when, as `true disks', we assume: \texttt{[case I]} isolated and paired disks; \texttt{[case II]} only isolated disks. In both the cases ongoing and post-coalescence mergers are considered as `true mergers.}
\label{well_corr_class}
\begin{tiny}
\centering
\hspace{-4mm}
\begin{tabular}{c | ccc | ccc}
\hline\hline\noalign{\smallskip}
Method & I$_{disk}$ (\%) & I$_{merger}$  (\%) & K$_{tot}$ &  I$_{disk}$ (\%) & I$_{merger}$ (\%)  & K$_{tot}$ \\
{\smallskip} 
(1) & (2) &(3) & (4)  & (5)  & (6)   & (7)   \\
&       &\texttt{[case I]}    &   &  &\texttt{[case II]} &  \\
\hline\noalign{\smallskip}
S08 (obs) low--z	&  25/34	(74)& 16/16 (100)   & 0.14 -- 0.15	& 10/13 (77) &16/16 (100)  &  0.13 -- 0.15\\
B12 (obs) low--z 	& 27/34 (79)	& 16/16 (100)   & 0.19		& 10/13 (77) &16/16 (100)  & 0.16 -- 0.19\\
\hline\noalign{\smallskip}
S08 (sim) high--z 	& 26/34 (76)	& 15/16 (94)   & 0.13 -- 0.14	& 11/13 (85) &15/16 (94)   & 0.1 -- 0.14\\
B12 (sim) high--z 	& 26/34 (76)	& 15/16 (94)   & 0.14 -- 0.17	& 11/13 (85) &15/16 (94)  & 0.11 -- 0.17\\
\hline\hline\noalign{\smallskip}
\end{tabular}
\vskip0.2cm\hskip0.0cm
\end{tiny}
\footnotesize
Col (1): Method used: S08 and B12 stands for `unweighted' and `weighted' methods. `Obs' and `sim' stand for observed and simulated samples. Col (2): Fraction of well classified disks over total number of disks as defined in Eq. 6. when isolated and paired disks are considered as `true disks'. The corresponding percentage is in brackets. Col (3): Fraction of well classified mergers over total number of mergers as defined in Eq. 6.  when isolated and paired disks are considered as `true disks'. The corresponding percentage is in brackets. Col (4): Value of the total (optimal) kinematic asymmetry derived when isolated and paired disks are considered as `true disks'. Col (5, 6, 7): Respectively, the same values as in Col (2, 3, 4) when only isolated disks are considered as `true disks'.
\end{table*}

\subsection{Comparison between low-- and high--z \texttt{kinemetry}--based results in the literature}

In this section we discuss the results obtained in this work with those derived in \citet{Hung15} ({hereafter, H15}) for an overlapped subsample of 8 interacting systems\footnote{The overlapped subsample is composed by the following galaxies: F06076--2139 (type 1), 08355--4944 (type 2), F10038--3338 (type 2), F10257--4339 (type 2), F12043--3140 (type 1), F17138--1017 (type 2), F18093--5744 (type 1), F23128--5919 (type 1).}. The H15 analysis is based on the application of the unweighted and weighted \texttt{kinemetry}--based methods to a sample of local (U)LIRGs observed with Wide Field Spectrograph (WiFeS) and artificially redshifted to z=1.5 degrading spatial resolution and sensitivity.

Two of these systems (F06076--2139 and F12043--3140) clearly show in the HST and DSS images, respectively, the presence of two merging galaxies in each system, which could not be resolved when simulated at high--z in H15. Their simulated kinematic maps show a complex and irregular pattern and are classified as `merger' according to the B12 criteria. Excluding these two systems, our kinematic classification of the 6 remaining systems (8 galaxies) is in good agreement with their findings. We only find disagreement for 2 of these galaxies (F10257--4339 and F18093--5744 S), classified as `merger' and `disk', respectively, in this work. Thus, for these 8 galaxies, the same disk/merger fraction (2/6) is derived in both the works according to the frontier considered in H15 and that derived in this analysis. If we consider the results derived for our simulated subsample at z=3 the derived disk/merger ratio becomes 3/5, since F17138--1017 shows more ordered kinematic maps at high--z than locally, classified as `disk' according to this analysis.

For this subsample the fractions of correctly identified disks and mergers according to our analysis are I$_{disk}$  = 2/3 and I$_{merger}$  = 5/5; according to the H15 analysis, I$_{disk}$  = 0/1, I$_{merger}$  = 5/7, since IRAS F17138--1017 is the only galaxy morphologically classified as isolated disk but kinematically classified as merger. Thus, a larger number of well classified disks and mergers is derived according to our analysis (67\% and 100\%) with respect to that derived in H15 (0\% and 71\%).

A smaller fraction of mergers at high--z has been also derived by \citet{Gon10}, observing a set of Lyman Break Analogs (LBAs) at z $\sim$ 0.2 and redshifted their sample at z = 2.2. The worse resolution of their simulated maps let decrease the fraction of mergers from low-- to high--z, respectively, from $\sim$ 70\% to $<$ 30\% according to the S08 limit. 
 
The angular resolution at which a sample is observed plays a key role in classifying galaxies as `disk' or `merger'. On the one hand, the loss of angular resolution, when simulating individual galaxies at high--z, tends to smooth the asymmetries in their kinematic maps, making objects to appear more `disky'; on the other hand, when simulating close interacting systems at high--z, it could result in unresolved systems which show more complicated kinematics than if resolved.

\subsection {The relationships between the K$_{tot}$ versus  L$_{IR}$, v$^*$/$\sigma$ and the projected nuclear separation}

In this section the relations between the total kinematic asymmetry K$_{tot}$ and some kinematical and dynamical parameters are considered. In particular, some trends are found when considering the K$_{tot}$ as a function of the infrared luminosity L$_{IR}$, the dynamical ratio v$^*$/$\sigma$\footnote{The v$^*$/$\sigma$ is the intrinsic dynamical ratio defined as the ratio of the intrinsic velocity shear to the mean velocity dispersion. See B13 for further details on how these parameters have been computed.} and the projected nuclear separation. Since at low--z the K$_{tot}$ values for the unweighted and weighted planes are only slightly different and the same general trend is conserved, we take into account the unweighted values (i.e., K$_{tot}$ = 0.145) for a possible comparison with other previous works (e.g., \citealt{Gon10, AZ12, swin12}).

\subsubsection{The K$_{tot}$ -- L$_{IR}$ relation}

\begin{figure}[htbp]
\centering
\includegraphics[width=0.45\textwidth]{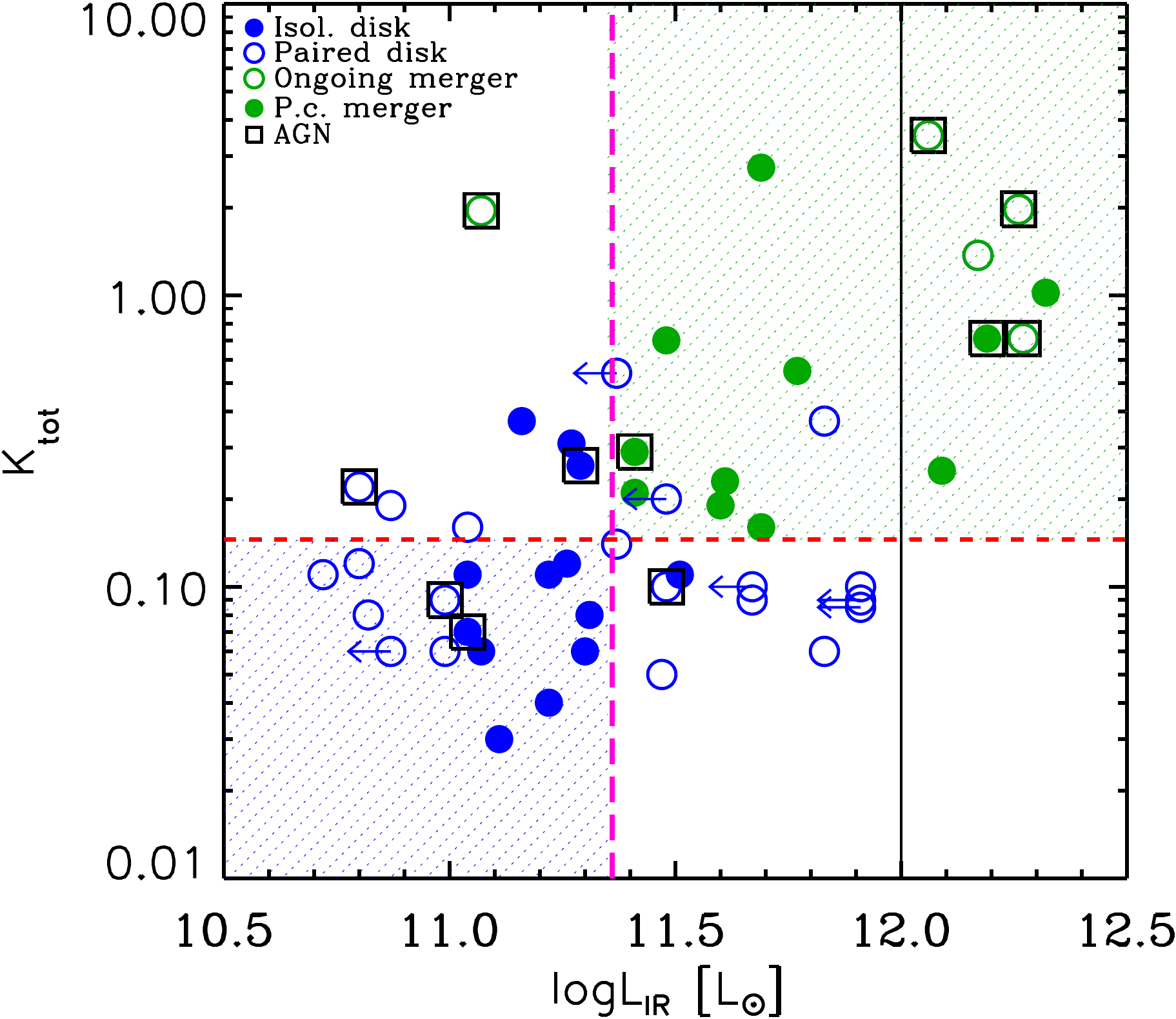}
\caption{Relation between the total kinematic asymmetry values, K$_{tot}$, as a function of the infrared luminosity, L$_{IR}$. The vertical solid black line separates the LIRG--ULIRG domain, while the vertical dashed pink line represents the infrared luminosity value which could separate `disks' from `mergers'. The horizontal dashed red line identifies the K$_{tot}$(I$_{max}$) = 0.145 value.}
\label{K_IR}
\end{figure}

In Fig. \ref{K_IR} the linear trend between the total kinematic asymmetries K$_{tot}$ as a function of the infrared luminosity L$_{IR}$ is shown. This plot clearly shows the (morphological and kinematical) results summarized in Tab. \ref{kinekine}. The majority of the objects with a luminosity L$_{IR} \ge$ 11.4 L$_\odot$ show high total kinematic asymmetries (19 out of 28 galaxies with K$_{tot}$ $>$ 0.14, green dashed area) and are classified as {\it mergers}. On the other hand, most of the less luminous ones (L$_{IR} <$ 11.4 L$_\odot$, 15 out of 22 galaxies with K$_{tot}$ $<$ 0.14, blue dashed area) have lower kinematic asymmetries and are classified as {\it disks}.  Thus, the luminosity value of log L$_{IR} \sim$ 11.4 L$_\odot$ seems to suggest it could be considered as a threshold value able to distinguish {\it disks} from {\it mergers}, but a sample complete in luminosity is needed to confirm this result.

A correlation between the morphology and the L$_{IR}$ has been already found in \citet{V02}: they derived that LIRGs are generally spirals which show a morphology much less disturbed than that shown in ULIRGs in the early phase of the interaction.

\subsubsection{The K$_{tot}$ -- v$^*$/$\sigma$ relation}

A clear correlation between the different phases of the merging process and the mean kinematic properties inferred from the kinematic maps has been found in our sample. In particular, isolated disks, interacting galaxies, and merging systems define a sequence of increasing mean velocity dispersion and decreasing velocity field amplitude, which is characterized by intrinsic average dynamical ratios (v$^*$/$\sigma$) of 4.7, 3.0, and 1.8, respectively (see B13).

\begin{figure}
\hskip-5mm\centering
\includegraphics[width=0.45\textwidth]{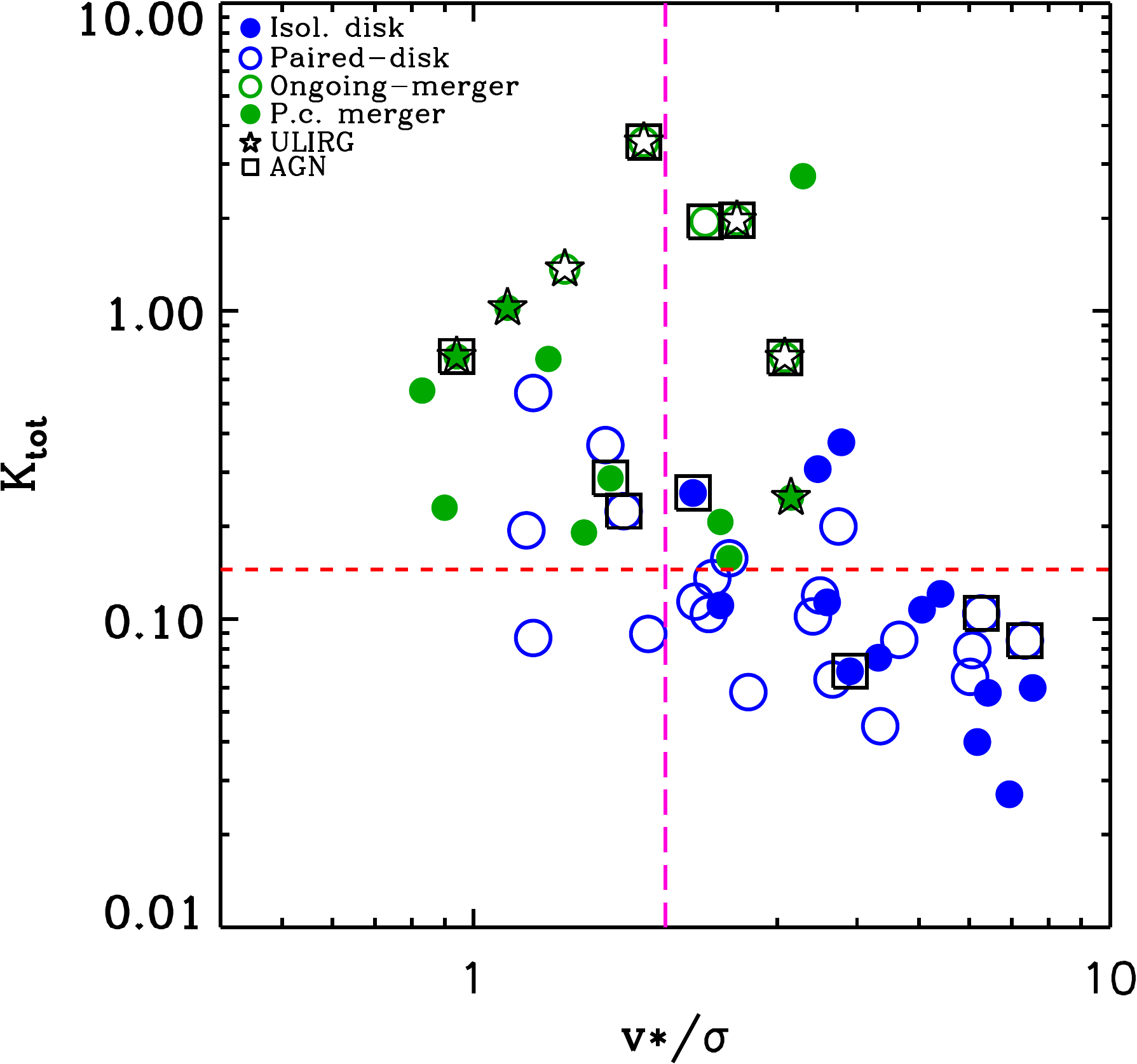}
\caption{The relation between the total kinematic asymmetry K$_{tot}$ and the intrinsic dynamical ratio v$^*$/$\sigma$ is shown. The more rotation dominated objects (v$^*$/$\sigma$$>$ 2) are those showing the lower kinematic asymmetry, while an opposite trend is found for the dispersion-dominated objects. The colors and symbols used are the same as in the previous figures. Black stars represent ULIRGs while black square identifies AGN. The horizontal dashed red line is the same than that shown in Fig. \ref{K_IR}. The vertical dashed pink line represents the value v$^*$/$\sigma$ = 2.}
\label{evolution}
\end{figure}

In a similar way, the total kinematic asymmetry K$_{tot}$ quantifies the kinematic asymmetry degree in a galaxy with respect to the ideal rotating disk case. In Fig. \ref{evolution} the relation between (unweighted) K$_{tot}$ and v$^*$/$\sigma$ for each source is considered. A (linear log -- log) inverse trend is found: the more rotation-dominated objects (v$^*$/$\sigma$ $>$ 2) generally show lower values of the total kinematic asymmetries (K$_{tot}$ $<$ 0.14) with respect to those derived for dispersion-dominated systems (i.e., v$^*$/$\sigma<$ 2, K$_{tot}$ $>$ 0.14), as expected. In order to quantify how well the 1D parameter v$^*$/$\sigma$ classifies disks and mergers in our sample with respect to the \texttt{kinemetry} results, we compute the fraction of well classified objects as before (i.e., Sect. \ref{distinction} and \ref{blabla_high}) deriving an index I = 1.4. This value is smaller than the one obtained using the total kinematic asymmetry K$_{tot}$, which is indicative that when the full 2D information is taken into account to study the kinematic asymmetries a better classification is obtained. All the ULIRGs are well classified as {\it mergers} according to our \texttt{kinemetry} frontier while only a small fraction (3 out of 7) is classified as such according to the v$^*$/$\sigma$ parameter. This confirms the importance of the 2D kinematic analysis in unveiling the real status of these systems. 

Our dynamical ratio threshold (v$^*$/$\sigma$ = 2) is in good agreement with that derived by \citet{Kassin12}. Indeed, studying the kinematics of a large sample of 544 blue galaxies over the last $\sim$ 8 billion years (0.2 $<$ z $<$ 1.2), they found that such systems become progressively more ordered with time as distorted motions decrease and rotation velocities increase. They define a kinematically `settled disk' as having a ratio of ordered/random motions larger than three (v/$\sigma$ $>$ 3), also deriving that the fraction of settled disks increases with time (decreases with z) since z = 1.2 for galaxies with stellar mass over 8 $<$ log M$_\star$ $<$ 10.7. The kinematic disk settling has be explained as due to: 1) a high frequency of merging at high--z and 2) higher gas fraction at early times. Since both these factors decrease with time, a general kinematic settling is expected with time (`kinematic downsizing'). According to their work, the galaxies settle to become the rotation-dominated disks found in the universe today, with the most massive galaxies being the most evolved at any time. Furthermore, at all redshifts they found that the most massive galaxies are on average the most kinematically settled while the least massive galaxies the least kinematically settled.

In our analysis we derived a similar trend: the most massive (log$<$M$_{dyn}$$>$ = 10.71 M$_\odot$, median value 10.69 M$_\odot$) and (morphologically) regular objects (class 0 isolated galaxies) show the highest dynamical ratio (v$^*$/$\sigma$ = 4.7) while the less massive pre- and post-coalescence galaxies (log$<$M$_{dyn}$$>$ = 10.68 M$_\odot$, median value 10.54 M$_\odot$ and 10.67 M$_\odot$, median value 10.23 M$_\odot$, respectively; see Tab.2 in B13) are characterized by lower v$^*$/$\sigma$, of 3.0 and 1.8, respectively. 

If we apply a threshold value of v$^*$/$\sigma$ = 3 to our data (see Fig. 5), it also suggests a good frontier to distinguish our systems in `disks' and `mergers'. In such a case, 21 out of 34 objects are well classified as `disks', while 13 out of 16 are well classified as `mergers', deriving an index parameter I = 1.43. This value is larger than the one derived when using the v$^*$/$\sigma$ = 2, but still lower with respect to that derived when using \texttt{kinemetry}, which gives the largest number of well classified `disks' and `mergers'.

\begin{figure}
\includegraphics[width=0.48\textwidth]{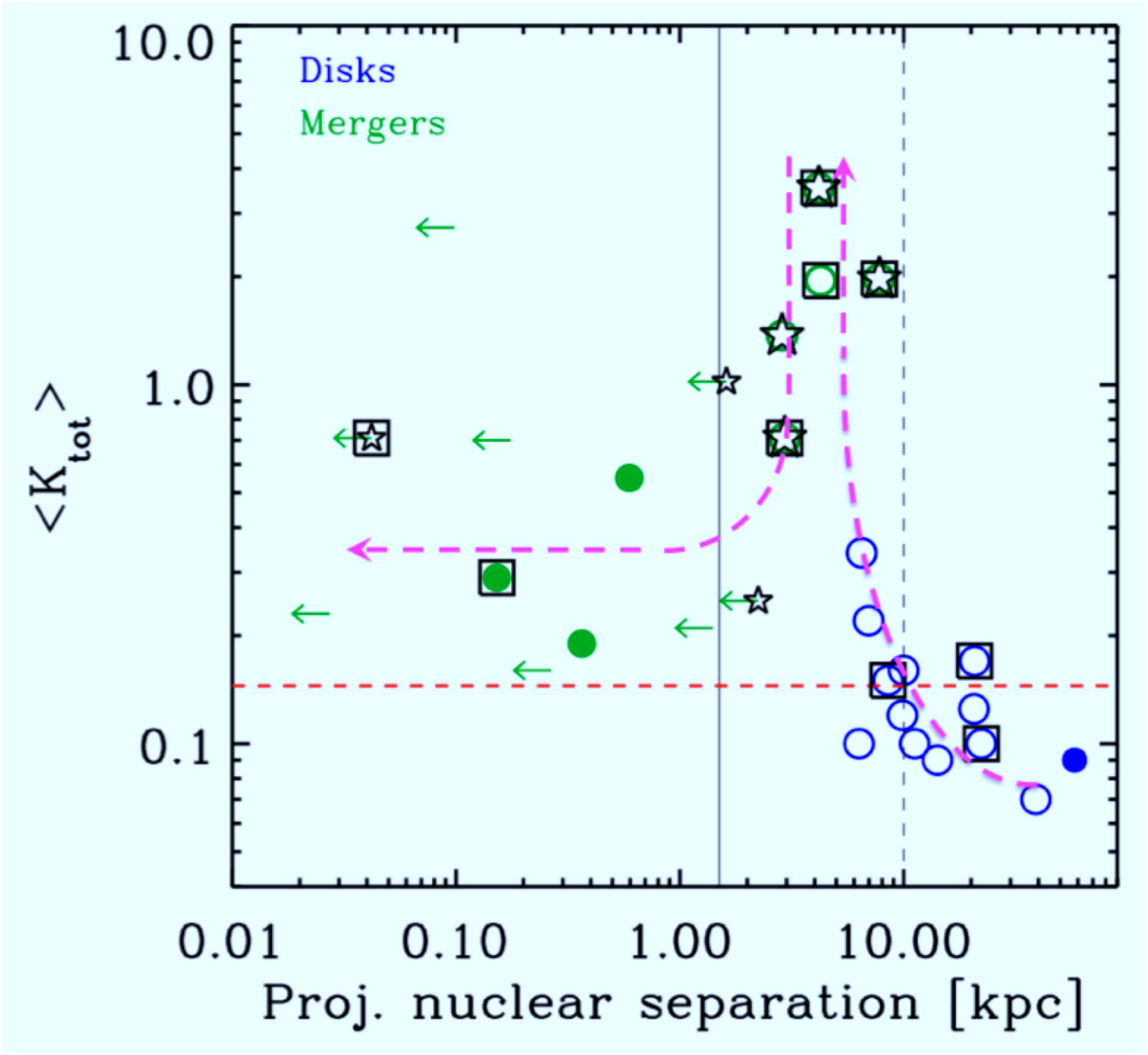}
\caption{Relation between the mean total kinematic asymmetry K$_{tot}$ and the nuclear projected separation in (wide and close) interacting and merging systems. The K$_{tot}$ of a system generally increases with the decreasing of the nuclear separation: it reaches its maximum value in ongoing (pre-coalescence) mergers, then decreasing for post-coalescence mergers. The colors and symbols used are the same as in the previous figure (Fig. \ref{evolution}). The pink dashed line helps to highlight the evolutionary trend found when a galaxy goes from the wide-interacting phase to the post-coalescence one. The horizontal dashed red line is the same than that shown in Fig. \ref{K_IR}. }
\label{evol}
\end{figure}

\subsubsection{Kinematic asymmetries as a function of the nuclear separation along the merger process}

We analyze the relation between the projected nuclear separation and the total kinematic asymmetry K$_{tot}$ for those pair of galaxies for which a nuclear separation can be computed. In particular, the nuclear separation can be estimated for 19 systems\footnote{We refer to the type 0 system IRAS F07027--6011, type 1 systems IRAS F01159--4443, IRAS F01341--3735, IRAS F06035--7102, IRAS F06076--2139, IRAS F06206--6315, IRAS F06259--4780, IRAS 08424--3130, IRAS F08520--6850, IRAS F09437+0317, IRAS F12043--3140, IRAS 12596--1529, IRAS F14544--4255, IRAS F18093--5744, IRAS F22491--1808, IRAS F23128--5919 and to a few type 2 objects IRAS 08355--4944, IRAS F10038--3338 and IRAS F21453--3511.}. An upper limit has been computed for the majority of the type 2 galaxies (assuming a nuclear separation smaller than the resolution element of the image considered). The HST and VIMOS continuum images have been used to derive the projected nuclear separation. In three cases (i.e., IRAS F01341--3735, IRAS F09437+0317, IRAS F14544--4255) the DSS images have been used since a larger FoV was needed to cover the whole system. 

In Fig. \ref{evol} the mean value of the (unweighted) K$_{tot}$ ($<$$K_{tot}$$>$) of each system is related to its nuclear projected separation. This plot highlights the fact that, during the first phases of the merging process, the smaller the nuclear separation of a system the higher its total kinematic asymmetry. Moreover, the maximum asymmetry value K$_{tot}$ is reached for the ongoing mergers with nuclear separation $\sim$ 2-5 kpc, in which the merger phase is currently taking place. Then, a more unclear trend is shown for the post-coalescence mergers (i.e., type 2; nuclear separation less than 1.5 kpc), although their values are generally lower than those characterizing ongoing mergers. The virialization of the inner parts of these objects explains such results. The pink dashed line in Fig. \ref{evol} helps to highlight the possible average evolutionary sequence obtained when a galaxy goes through the wide-interacting to the post-coalescence phases. 

As shown in this plot, a trend is found between the interaction stage (based on a morphological classification) and the kinematic asymmetries of our systems. Our results are in agreement with those presented in H15 (Fig.~3 in their work), where the fraction of disk/merger galaxies classified using the \texttt{kinemetry} criteria is shown as a function of the interaction stage. The agreement between our and their results can be explained as follows. A galaxy with a low value of the total kinematic asymmetry K$_{tot}$ ($<$0.16) results in a `disk'-like (regular) kinematics while a high K$_{tot}$ value ($>$0.9) corresponds to a more disturbed and complex kinematics. Thus, K$_{tot}$ can be considered as a proxy of the fraction of mergers. In their work, going through the different interaction stages, from isolated to post-coalescence objects (from S to M4 stages), the maximum merger fraction is reached in the case of merged galaxies which still show two distinct nuclei (M3): their M3 galaxies correspond to our close-interacting systems, that show the highest K$_{tot}$ values. In particular, the mean K$_{tot}$ values (see Tab.~\ref{kinekine}) as a function of the interaction stage (isolated and paired disks, ongoing- and post-coalescence mergers) reproduce a trend similar to that derived by H15.

\section{Summary}

We have carried out a \texttt{kinemetry}-based classification for a large sample of 38 local (z $<$ 0.1) (U)LIRG systems observed with VIMOS/VLT with IFS. The sample covers a wide range of morphological types (i.e., spirals, interacting systems and merger remnants) and it is therefore well suited to study how the \texttt{kinemetry}--based criteria are able to distinguish {\it disks} and {\it mergers} in our objects. The disk/merger fraction allow us to constrain different evolutionary scenarios. Specifically we have applied the S08 (`unweighted') and B12 (`weighted') criteria to derive the total kinematic asymmetry (K$_{tot}$) to our observed sample as well as to simulated data, `redshifting' our sample at z=3. From our analysis we draw the following conclusions:

\vspace{3mm}

{\bf Kinematic distinction between {\it disks/mergers}}

\begin{enumerate}

\item The kinematic properties derived using the \texttt{kinemetry}--based methods are consistent with their morphological classification. The results obtained using the weighted and unweighted methods are similar when the whole sample is considered. 
\vspace{3mm}

\item

We can distinguish our sample in three kinematic groups according to the total kinematic asymmetry value K$_{tot}$ when using the weighted (unweighted) method: 1) 25 out of 50 galaxies are kinematically classified as `disk', with a K$_{tot}$ $\leq$ 0.16 (0.14); 2) 9 out of 50 galaxies are kinematically classified as `merger', with a K$_{tot}$ $\geq$ 0.94 (0.66); 3) 16 out of 50 galaxies lie in the `transition region', in which `disks' and `mergers' coexist, with 0.16 (0.14) $<$ K$_{tot}$ $<$ 0.94 (0.66).

\vspace{3mm}

\item
The K$_{tot}$ frontier value that better classifies the highest numbers of `disks' and `mergers', according to the morphology is K$_{tot}$ = 0.19 ($\sim$0.15): we obtain 27 (25) `disks' and 23 (25) `mergers' according to this value. 

The percentages of `correctly identified' disks and mergers at low--z result in, respectively, 79\% (74\%) -- 100\%. If only isolated disk galaxies are considered as `true disks', similar fractions are obtained with both the methods.

\vspace{2mm}
\item
When we apply our criteria to our systems simulated at z=3 just considering resolution effects, a lower total kinematic asymmetry frontier (K$_{tot}$ $\sim$ 0.16 ($\sim$0.14)) with respect to that found locally is derived when using the weighted (unweighted) method. We obtain 26 `disks' and 24 `mergers' according to this value. 

However, the `correctly identified' disks and merger fractions for the simulated high--z objects is 76\% -- 94\% with both the methods. If only isolated disk galaxies are considered as `true disks', these values become, respectively, 85\% and 94\%.

The loss of angular resolution makes objects to appear more kinematically regular (`disky') than actually they are as a consequence of the smearing of the kinematic features.

\vspace{4mm}
{\bf Relationships between the \texttt{kinemetry}--based K$_{tot}$ and morpho--kinematic parameters}
\vspace{2mm}

\item 
A trend is found between the K$_{tot}$ and the infrared luminosity L$_{IR}$, with the most luminous objects (ULIRGs) showing the highest total kinematic asymmetries. Furthermore, the luminosity value log L$_{IR} \sim$ 11.4 L$_\odot$ suggests that it could be considered as a threshold value able to separate these two morphological classes. To confirm this, a sample complete in luminosity is needed. 

\vspace{2mm}
 
\item 
An inverse trend is derived between the K$_{tot}$ and the intrinsic dynamical ratio v$^*$/$\sigma$: morphologically classified {\it disks} show higher dynamical ratio (v$^*$/$\sigma$ $>$ 2) and lower total kinematic asymmetry K$_{tot}$ ($<$ 0.14). Contrary, for the {\it mergers} v$^*$/$\sigma$ is lower ($<$ 2) while K$_{tot}$ is higher ($>$ 0.14). 

Our results support the `kinematic downsizing' scenario proposed by \citet{Kassin12}, where systems become progressively more ordered with time as distorted motions decrease and rotation velocities increase, where the most massive galaxies are on average the most kinematically settled.
\vspace{2mm}

\item
An interesting trend is also found between the K$_{tot}$ and the projected nuclear separation (as a proxy of the galaxy interaction stage) along the merger process. The smaller the nuclear separation the larger the K$_{tot}$, which reaches its maximum value during the `ongoing merging phase' (nuclear separation between the galaxies of 2--5 kpc) and then decreases during the post-coalescence merging phase, although with a relatively large dispersion. Our results are in agreement with those derived in \citet{Hung15}, who found that the merger fraction (as a proxy of the K$_{tot}$ parameter) shows a strong trend with the galaxy interaction stage.

\vspace{4mm}

{\bf The robustness of the K$_{tot}$ frontier determination in classifying \texttt{disks} and \texttt{mergers}}
\vspace{2mm}

\item 

From our results, the kinematic frontier we derive to distinguish `disks' from `mergers' is well determined. Indeed, when type 1 (interacting) objects are included (case I) or not (case II) in the `disk' group, the derived fractions of `well classified' \texttt{disks} and \texttt{merger} in both the cases are very akin, $\sim$80\% and $\sim$100\%, respectively. \\ This result can also confirm that the `paired disk' objects can actually be considered as `disks' according to their kinematic asymmetries.  
 
\vspace{2mm}
\item 
The K$_{tot}$ limit derived by \citet{S08} (K$_{tot}$ = 0.5) to separate `disks' from `mergers' at high--z is $\sim$~65\% larger, respectively, than the one found by us for the whole sample (observed locally and simulated at high--z) with both the weighted and unweighted methods. The use of this frontier would imply that the number of `disks' in our sample would be largely overestimated (classifying the 85\% of the galaxies as `disk'), since only the ongoing-- and some of the post--coalescence mergers with more complex kinematics would be classified as `mergers'. This, together with the effects of resolution on high--z samples, suggests that the fraction of disks at high--z inferred from similar kinematic criteria may be overestimated. 

\vspace{2mm}

\item  

The value of the frontier derived using the \texttt{kinemetry}--based methods strongly depends on the morphological classification which is key when analyzing high--z SGFs. Such systems may be dominated by several mechanisms and characterized by different gas and dust content, stellar mass and interaction stage. Thus, the combination of high resolution morphology (such as those coming from HST or AO--assisted imaging) along with spatially resolved kinematics will allow one to reveal the dynamical state of such systems (\citealt{NEICHEL2008}). Multi-wavelength morphological observations are needed to study the molecular gas phase (closely related to star formation) as well as the stars to better constrain the disk/merger fraction of SFGs at high--z.

\end{enumerate}

\begin{acknowledgements}

We acknowledge the anonymous referee for useful comments and suggestions, that helped us to improve the quality and presentation of the paper. This work was funded in part by the Marie Curie Initial Training Network ELIXIR of the European Commission under contract PITN-GA-2008-214227. This work has been supported by the Spanish Ministry of Science and Innovation (MICINN) under grant ESP2007-65475-C02-01, PNAYA2010-21161-C02-01, AYA2010-21697-C05-01, PNAYA2012-32295, PNAYA2012-39408-C02-01. Based on observations carried out at the European Southern observatory, Paranal (Chile), Programs 076.B-0479(A), 078.B-0072(A) and 081.B-0108(A). This research made use of the NASA/IPAC Extragalactic Database (NED), which is operated by the Jet Propulsion Laboratory, California Institute of Technology, under contract with the National Aeronautic and Space Administration. 
\end{acknowledgements}

\bibliographystyle{aa} 
%\bibliography{biblio}     
%-------------------------------------------------------------

\end{document}